\documentclass[useAMS,usenatbib]{mn2e}
\usepackage{graphicx}
\usepackage{subfigure}
\usepackage{amsmath}
\usepackage{amssymb}
\def\nbody{$N$-body\ }
\def\Nbody{$N$-body\ }

\title[ Gravothermal oscillations ]{Gravothermal oscillations in two-component
models of star clusters}
\author[P. G. Breen and D. C. Heggie]{ Philip G. Breen$^1$\thanks{E-mail:
p.g.breen@sms.ed.ac.uk} and Douglas C. Heggie$^1$\thanks{E-mail:  d.c.heggie@ed.ac.uk} \\
$^1$ School of Mathematics and Maxwell Institute for Mathematical Sciences, University of Edinburgh, King’s Buildings, Edinburgh EH9 3JZ}

\begin{document}

\date{   \today }

\pagerange{\pageref{firstpage}--\pageref{lastpage}} \pubyear{2010}

\maketitle

\label{firstpage}

\begin{abstract}

In this paper, gravothermal oscillations are investigated in two-component clusters with a range of different stellar mass ratios and total component mass ratios. The critical number of stars at which gravothermal oscillations first appeared is found using a gas code. The nature of the oscillations is investigated and it is shown that the oscillations can be understood by focusing on the behaviour of the heavier component, because of mass segregation. It is argued that, during each oscillation, the re-collapse of the cluster begins at larger radii while the core is still expanding. This re-collapse can halt and reverse a gravothermally driven expansion. This material outside the core contracts because it is losing energy both to the cool expanding core and to the material at larger radii. The core collapse times for each model are also found and discussed. For an appropriately chosen case, direct $N$-body runs were carried out, in order to check the results obtained from the gas model, including evidence of the gravothermal nature of the oscillations and the temperature inversion that drives the expansion.

\end{abstract}


\section{Introduction}
Gravothermal oscillations are one of the most interesting phenomena which may arise in the post-collapse evolution of a star cluster. The inner regions of a post collapse cluster are approximately isothermal and are subject to a similar instability as the one found in an isothermal sphere in a spherical container, as studied by \cite{Antonov} and \cite{LyndenBellWood1968}. Gravothermal oscillations, which are thought to be a manifestation of this instability, were discovered by \citet{betsug1984} whilst studying the post-collapse evolution of star clusters using a gas model. For a gas model of a one-component cluster it was found that gravothermal oscillations first appear when the number of stars $N$ is greater than 7000 \citep{Goodman1987}. This value of $N$ has also been found with Fokker-Planck calculations \citep{Cohn_et_al1989} and by direct $N$-body simulations \citep{Makino1996}. However, in a multi-component cluster the situation is more complicated. The presence of different mass components introduces different dynamical processes to the system such as mass stratification. Multi-component systems try to achieve kinetic energy equipartition between the components, which causes the heavier stars to move more slowly and sink towards the centre. This can lead to the Spitzer instability \citep{Spitzer} in which the heavier stars continuously lose energy to the lighter stars without ever being able to reach equipartition. \citet{Murphyetal1990} found that the post-collapse evolution for multi-component models was stable to much higher values of $N$ than in the case of the one-component system and that the value of $N$ at which gravothermal oscillations appeared varied with different mass functions.

In order to gain a deeper understanding of gravothermal oscillations, it is desirable to work with simpler models in which some of the effects which are present in real star clusters are ignored or simplified. For example, real star clusters have a range of stellar masses present, but in the current paper, the stellar masses are limited to two. Gaseous models are often used in this kind of research \citep{betsug1984, Goodman1987, HeffieAarseth1992} because they are computationally efficient. \cite{KimLeeGood1998} have already completed research in this area using Fokker-Planck models. However, their research was limited to mostly Spitzer stable models and only a small range of stellar mass ratios. The study in the present paper looks at the more general Spitzer unstable models using various stellar mass and total mass ratios.

There is also evidence of gravothermal oscillations in real star clusters. \cite{GierszHeggie} modelled the cluster NGC 6397 using Monte Carlo models and found fluctuation in the core radius. Their timescale suggests that they are gravothermal. Subsequently, they confirmed these fluctuations using direct $N$-body methods with initial conditions generated from the Monte Carlo model \citep{HeggieGiersz}. 

Two-component clusters may seem very unrealistic but there is reason to believe that they may be a good approximation to multi-component systems. \cite{KimLee1997} were able to find good approximate matches for half-mass radius $r_h$, central velocity dispersion $v_c$, core density $\rho_c$ and core collapse time $t_{cc}$ between two-component models and eleven-component models which were designed to approximate a power law IMF. Also see \cite{KimLeeGood1998} for a discussion of the realism of two-component models.

This paper is structured as follows. In Section 2, we describe the models which are used. This is followed by Section 3, in which the results concerning gravothermal oscillations are given. Section 4 is concerned with the results of the core collapse times. In Section 5, the results of $N$-body simulations are given.  Finally Section 6 consists of the conclusions and a discussion. 

\section{Models}\label{sec:gaseq}

\subsection{Gas model}\label{sec:gasmodel}
\subsubsection{Basic equations and Notation}
In our model, we ignore primordial binaries and stellar evolution, and assume that the energy generating mechanism is the formation of binary stars in three body encounters and subsequent encounters of binaries with single stars. In a one-component model the rate of energy generation per unit mass is approximately
\begin{equation} \epsilon = 85\frac{G^5 m^5 n^2}{\sigma^7_c}\label{eq:energygenerationrate} \end{equation} \citep{HeggieHut2003}, where $m$ is the stellar mass, $n$ is the number density, $\sigma_c$ is the one dimensional velocity dispersion of the core and $G$ is the gravitational constant. 
\citet{Goodman1987}, whose results on the 1-component model we shall occasionally refer to, used a similar formula, with a coefficient which is, in effect, in the range 140--170 (depending on the value of $N$).

The equations of the two-component gas model \citep{HeffieAarseth1992} are given below: 
\begin{eqnarray}\label{eq:one}
 \frac{\partial M_i}{\partial r} = 4 \pi \rho_i r^2 
\end{eqnarray}
\begin{eqnarray}\label{eq:two}
 \frac{\partial p_i}{\partial r} = - \frac{G (M_1+M_2)}{r^2} \rho_i 
\end{eqnarray}
\begin{eqnarray}\label{eq:three}
 \frac{\partial \sigma_i}{\partial r} = - \frac{\sigma_i L_i}{12\pi C m_i \rho_i r^2 \ln \Lambda}
\end{eqnarray}
\begin{eqnarray}\label{eq:four}
\frac{\partial L_i}{\partial r} = & -4 \pi r^2 \rho_i \Big[ \sigma^2_i \Big( \dfrac{D}{Dt} \Big)  \ln{\Big(\dfrac{\sigma^3_i}{\rho_i} \Big)}  +  \delta_{i,2}\epsilon  \\ 
& + 4 (2 \pi)^{\frac{1}{2}} G^2 \ln{\Lambda} \Big[ \dfrac{\rho_{3-i}}{(\sigma^2_1+\sigma^2_2)^{\frac{3}{2}}} \Big] ( m_{3-i} \sigma^2_{3-i} - m_i \sigma^2_i )  \Big]  \nonumber
\end{eqnarray}
where $i=1,2$. This model in turn is ultimately inspired  by the one-component model of \cite{LyndenBellEggleton1980}.

\begin{table}
 \caption{ Notation $($the subscript $i$ corresponds to the $i^{th}$  component, $i=2$ refers to the more massive component $)$}
\begin{tabular}{ |c|c| }
\hline 
$r$ & radius \\
$\rho$ & mass density 	\\ 
$\sigma $    & one dimensional velocity dispersion 	\\ 
$m$  & stellar mass  \\
$M$  & total mass (within radius r)   \\
$C$  & Constant (see text)\\
$L$  & energy flux \\
$N$  & number of stars \\ 
$ \ln{\Lambda}$ & coulomb logarithm ($\Lambda=0.02N$) \\
$\dfrac{D}{Dt}$& Lagrangian derivative (at fixed $M$)\\
 $\dfrac{\partial}{\partial r}$& radial derivative (at fixed $t$)\\
 \hline
\label{table:tabterms}
\end{tabular}

\end{table}

The meaning of the symbols can be found in Table \ref{table:tabterms}. The major difference between the above equations and those for the one-component model is the last term of equation \ref{eq:four}, which involves the exchange of kinetic energy between the two components. See Spitzer (1987, p.39) for information on this term. As the heavier component dominates in the core of the cluster, it is assumed that all of the energy is that generated from the second component. Hence the Kronecker delta $\delta_{2,i}$ in the last equation. There are two constants in the gas code which can be adjusted: $C$ and the coefficient $\lambda$ of $N$ in $\Lambda=\lambda N$. The value of $\lambda=0.02$ was used as it was found to provide a good fit for multi-component models \citep{GierszHeggie2}. The value of $C$ used was $0.104$ \citep{Heggieramamani}. This value of $C$ results from the comparison of core collapse between gas and Fokker-Planck models of single component systems and it is not clear if it applies accurately to post-collapse two-component models.
\subsubsection{The role of $N$ in the gas code}

This paper places emphasis on the role of $N$ in evolution, but it is not clear what role $N$ plays in equations (\ref{eq:one}) $\mbox{--}$ (\ref{eq:four}). For fixed structure (i.e. $\rho_i(r),$ etc), $N$ appears explicitly in $\Lambda$ (where its role is rather insignificant), and in the individual masses $m_i$. These appear in equations (\ref{eq:three}) and (\ref{eq:four}). In a system with fixed structure, equation (\ref{eq:three}) shows 
$$L_i \propto m \ln{\Lambda} \propto \frac{\ln{\lambda} N}{N} ,$$
reflecting the fact that the flux $L$ is caused by two-body relaxation, and its time scale is proportional to $\frac{N}{\ln{\lambda N}}$. In equation (\ref{eq:four}) $N$ plays a similar role in the last term on the right, which governs the approach to equipartition. It also appears implicitly through $\epsilon$, because of the $m$ dependence in equation (\ref{eq:energygenerationrate}). For a system of given structure, its contribution to $L$ in equation (\ref{eq:four}) is proportional to $N^{-3}$ (as we are assuming that $\rho=mn$ is fixed and so $\epsilon$ in equation (\ref{eq:energygenerationrate}) is proportional to $m^3$). It would seem as though this term is insignificant for large $N$. In practice, however, the system compensates by increasing the central density so that $\epsilon$  plays a comparable role to the relaxation terms (see Section \ref{sec:DOH}).

\subsection{Direct \Nbody}
Direct \Nbody simulations were conducted using the NBODY6 code (Aarseth, 2003) enabled for use with Graphical Processing Units (GPU). NBODY6 has a range of features and options such as individual time steps which make it an excellent direct \Nbody code. NBODY6 is written in FORTRAN and is publicly available for download from www.ast.cam.ac.uk/$\!\sim\,$sverre/web/pages/nbody.htm\ .

\section{ Critical Value of $N$ }\label{sec:Ncritinto}
\begin{figure}
\subfigure{\scalebox{0.45}{\includegraphics{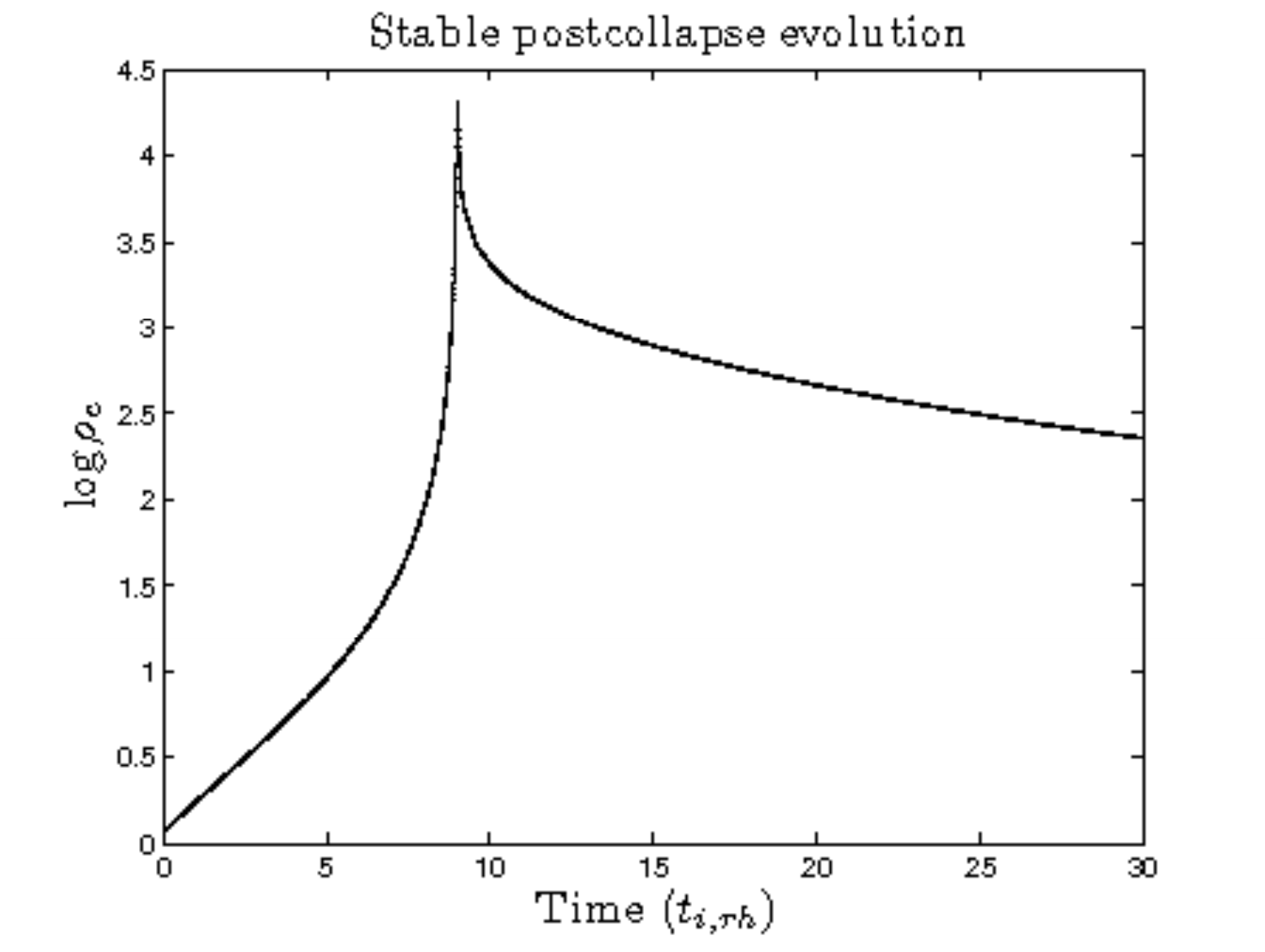}}}\quad
\subfigure{\scalebox{0.45}{\includegraphics{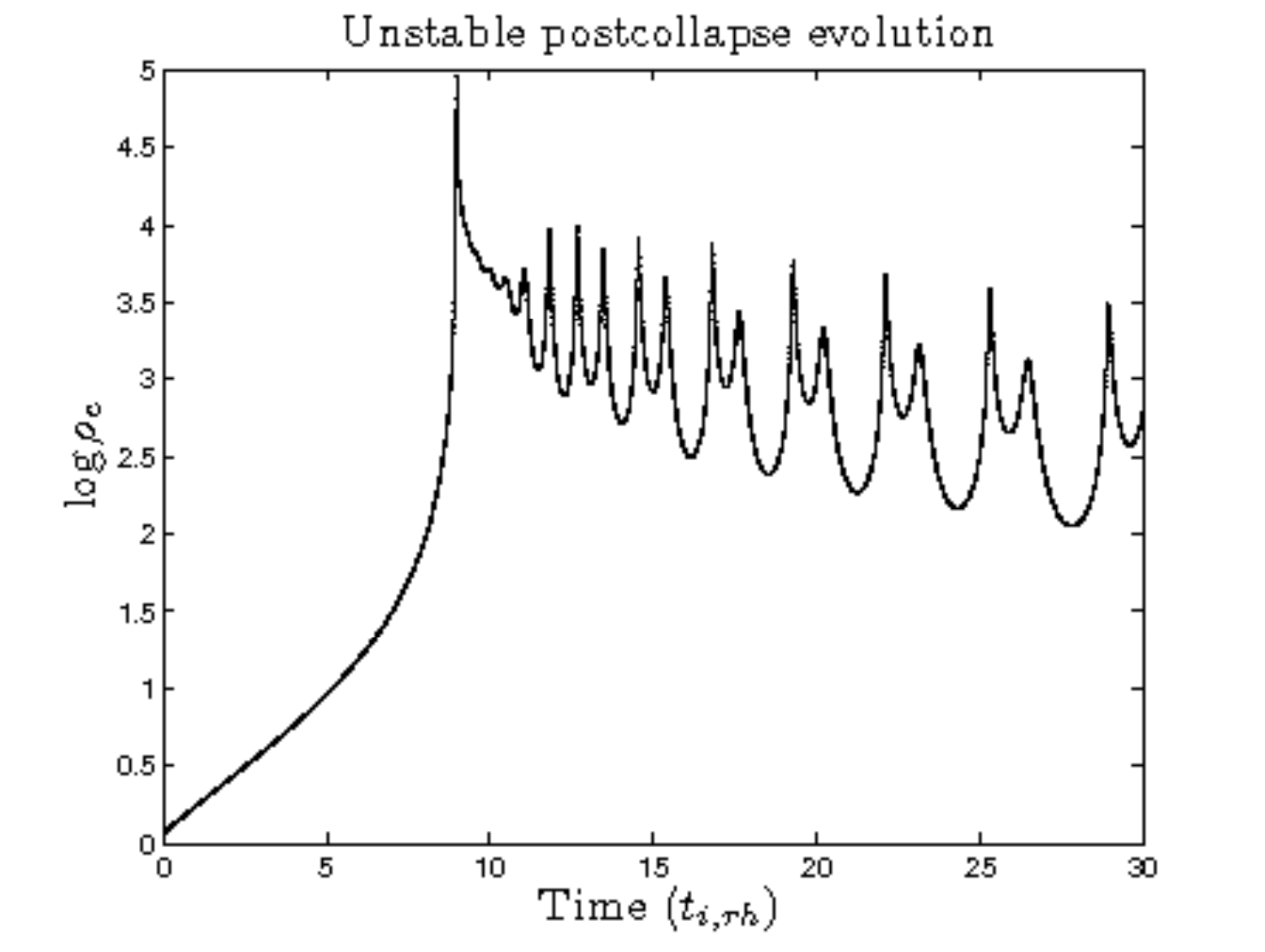}}}

\caption{Logarithm of the central density vs time (in units of the initial value of $t_{rh}$) for a two-component gas model, $\frac{m_2}{m_1}=2$, $\frac{M_2}{M_1}=1$, top: $N = 1.5 \times 10^4$ (stable), bottom: $2.5 \times 10^4$ (unstable). For initial conditions see Section \ref{sec:Ncrit} }
\label{postcollcomp}
\end{figure}

If the value of $N$ is not too large, then, after core collapse, the cluster expands at a steady rate (Fig. \ref{postcollcomp}, top). However, at a larger value of $N$ the central density $(\rho_c)$ was found to oscillate (Fig. \ref{postcollcomp}, bottom). \cite{Goodman1987} showed that for one-component models  the steady expansion is unstable for large values of $N$ and found that the value at which oscillations first appeared is $N=7000$. In this present paper, the case of two-component models is investigated.  

\subsection{Results of the gas code}\label{sec:Ncrit}
\begin{figure}
\includegraphics[scale=0.55]{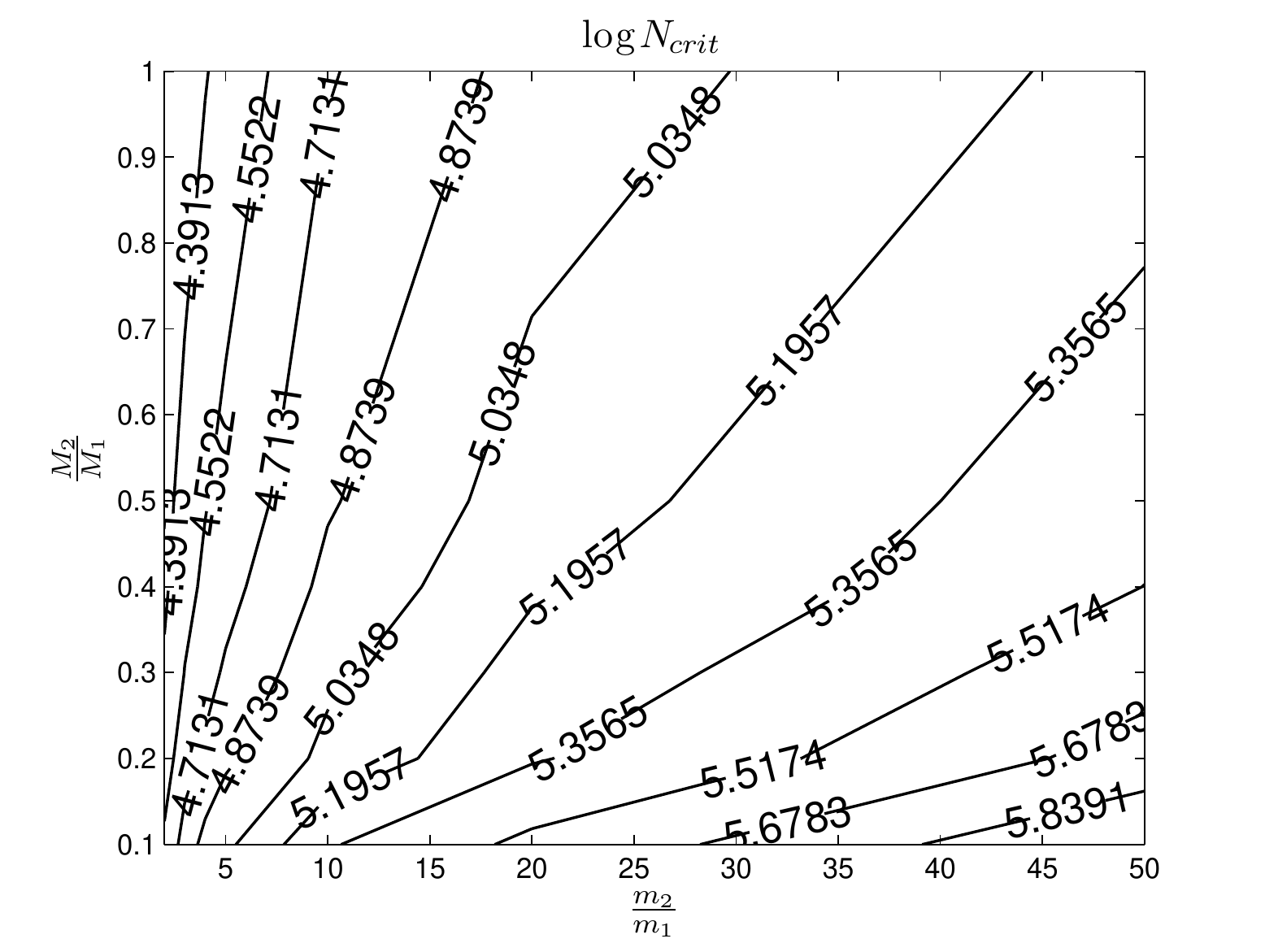}
\caption{Contours of $\log_{10}(N_{crit})$}
\label{logNcirt}
\end{figure}

In all cases, the initial conditions used were Plummer models \citep{Plummer1911, HeggieHut2003}. The initial velocity dispersions of both components were equal and the initial ratio of density of each component was equal at all locations. The initial conditions were constructed with different stellar mass ratios $\frac{m_2}{m_1} = 2,3,4,5,10,20,50$ and for each of these mass ratios, a model with total mass ratios $\frac{M_2}{M_1}=0.1,0.2,0.3,0.4,0.5$ and $1$ was constructed. A python script was used to run the gas model code over a range of values of $N$ for each of the pairs of mass ratios. Each run terminated when the time value reached 30 initial relaxation times ($t_{i,rh}$).The value of the central density was checked for an increase in value of 5 percent or more in any interval over the time period between $20t_{i,rh}$ and $30t_{i,rh}$. If an increase was found, the run was deemed to be unstable and the range of $N$ was refined. This process continued until the critical value of $N$ ($N_{crit}$) at which oscillations first appeared was determined (correct to ten percent). The values of $N_{crit}$ were also visually confirmed from the output of the gas code. The obtained values of $N_{crit}$ in units of $10^4$ are given in Table \ref{table:tab1}. Fig. \ref{logNcirt} shows a contour plot of $ \log_{10} N_{crit}$. 

\begin{table}
\begin{center}
\caption{Critical value of $N$ ($N_{crit}$) in units of $10^4$ }
\begin{tabular}{ c  |c |c| c |c |c| c|c|c|}
\cline{2-9} 
                    & \bf{1.0} & 1.7 & 2.0 & 2.4 & 2.8 & 5.0 & 8.5 & 18 	\\ \cline{2-9}
		    & \bf{0.5} & 2.2 & 2.8 & 3.5 & 4.0 & 7.2 & 13  & 30 	\\ \cline{2-9}
 $\frac{M_2}{M_1} $ & \bf{0.4} & 2.3 & 3.2 & 3.8 & 4.6 & 8.2 & 15  & 33  	\\ \cline{2-9}
                    & \bf{0.3} & 2.6 & 3.6 & 4.6 & 5.4 & 10  & 18  & 42		\\ \cline{2-9}
                    & \bf{0.2} & 3.0 & 4.4 & 5.5 & 7.0 & 12  & 22  & 55		\\ \cline{2-9}
                    & \bf{0.1} & 3.8 & 6.0 & 8.5 & 10  & 22  & 36  & 100	\\ \cline{2-9}
\cline{2-9}
                    &     & \bf{2}   & \bf{3}   & \bf{4}   & \bf{5}   & \bf{10}  & \bf{20}  & \bf{50} 	\\
\cline{2-9}
\label{table:tab1}
\end{tabular}
\\
 $\frac{m_2}{m_1}$ 
\end{center}
\end{table}

\subsection{Interpretation of the results}\label{sec:lsm}

\begin{figure}
\subfigure{\scalebox{0.45}{\includegraphics{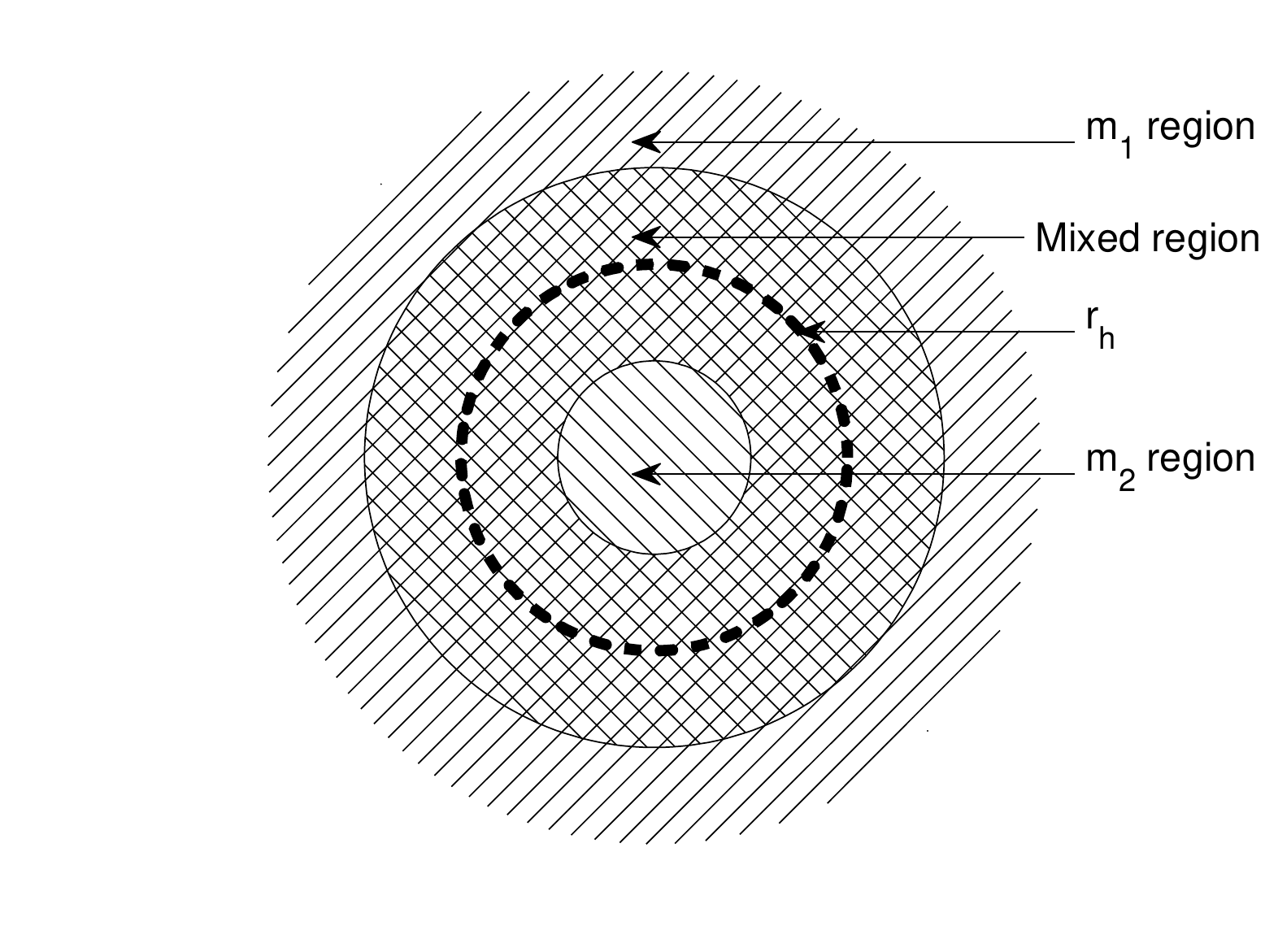}}}\quad

\caption{Illustration of mass distribution in star clusters.
 The dashed line represents $r_h$, the lines at $135$ degrees in the centre represent the area
dominated by the heavy component (i.e $\frac{\rho_2}{\rho_1}\gg 1 $), the lines
at $45$ degrees in the far halo represent the area dominated by the light component
(i.e $\frac{\rho_2}{\rho_1}\ll 1 $), the crossed section represents the area
where there is a mixture of heavy and light components (i.e.
$\frac{\rho_2}{\rho_1} \sim 1 $ ) }
\label{cartoon}
\end{figure}

In order to attempt to interpret the results in the previous subsection,
it is helpful to illustrate the mass density distribution of each component
within the cluster and this is done in Fig. \ref{cartoon}. 

\begin{figure}
\subfigure{\scalebox{0.26}{\includegraphics{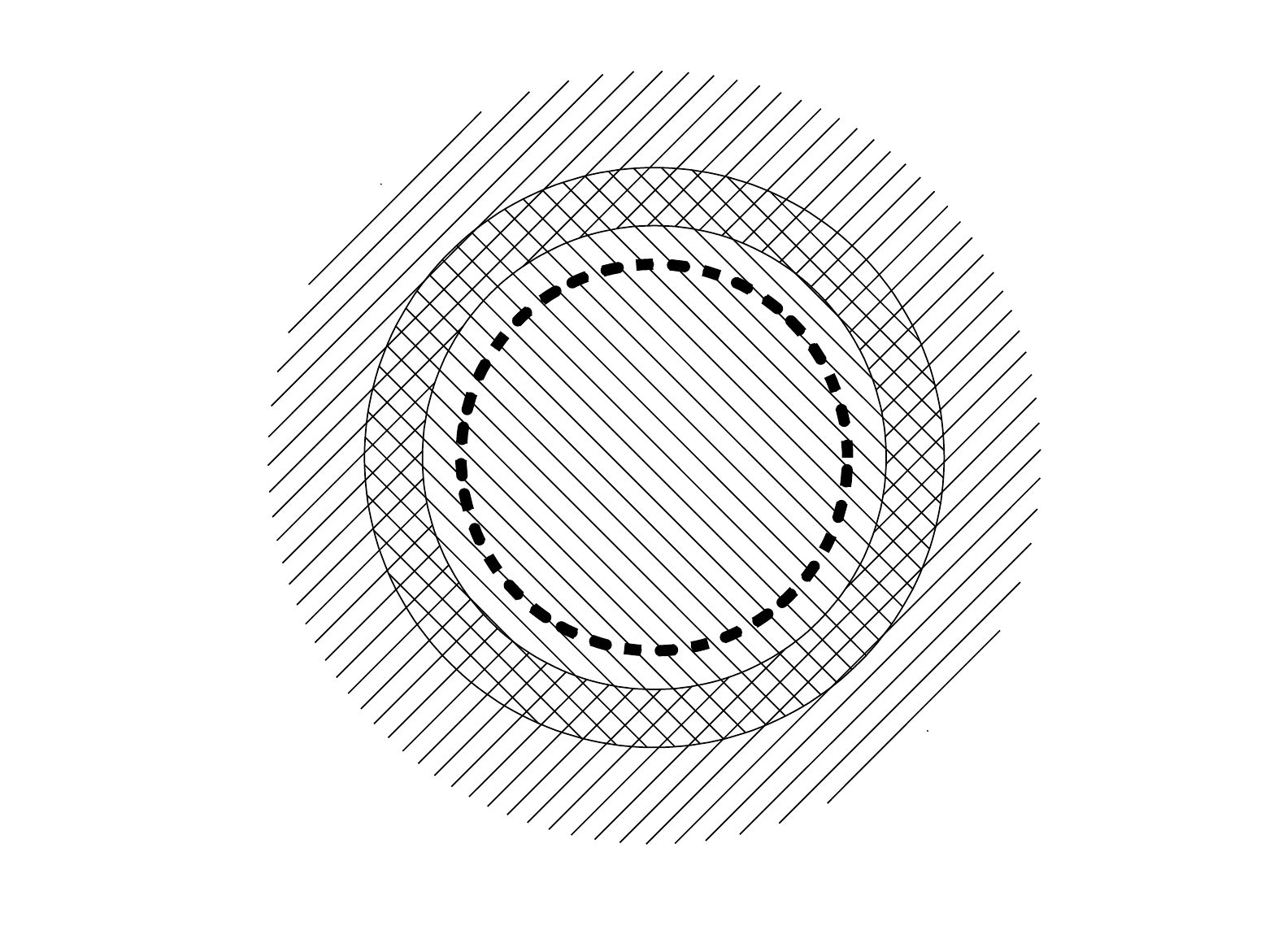}}}
\subfigure{\scalebox{0.26}{\includegraphics{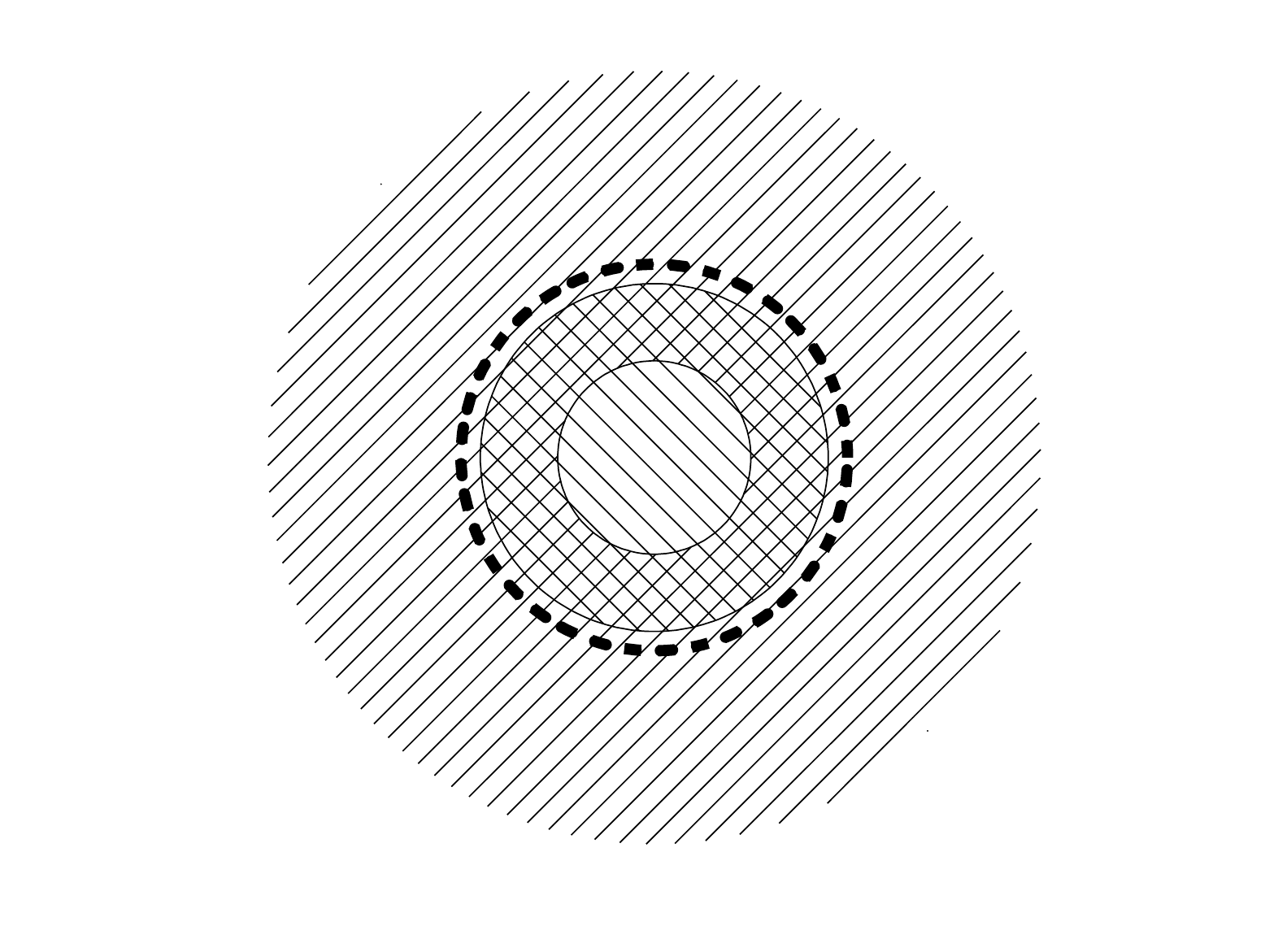}}}

\caption{ Left: System with $\frac{M_2}{M_1} \gtrsim 1$ and with large enough
$\frac{m_2}{m_1}$ to remove most of the light component from within the half mass
radius. Right: System with $ \frac{M_2}{M_1} < 1$ and $\frac{m_2}{m_1} \gg 1$. }
\label{cartoon2}
\end{figure}

\begin{figure}
\subfigure{\scalebox{0.25}{\includegraphics{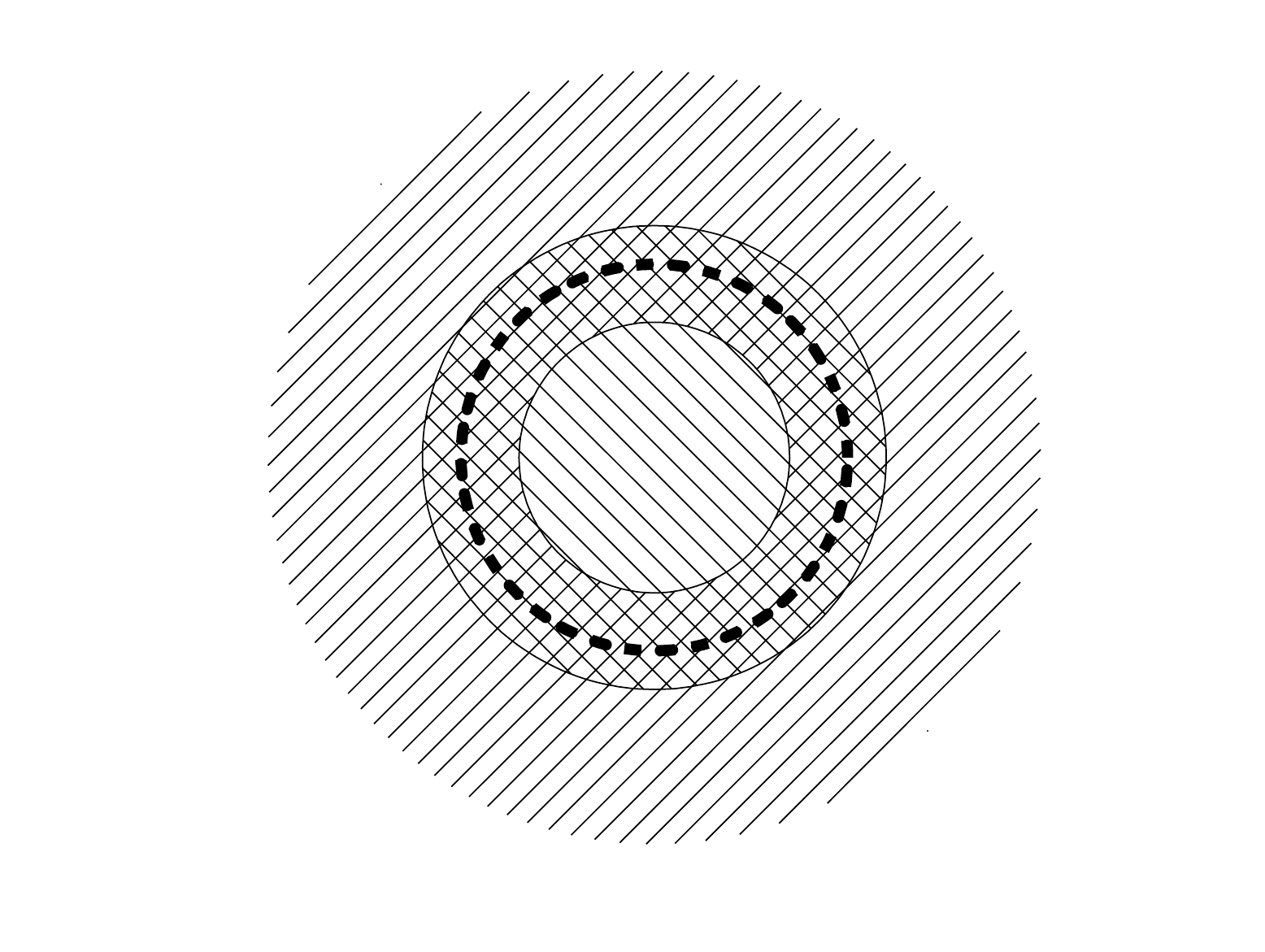}}}
\subfigure{\scalebox{0.25}{\includegraphics{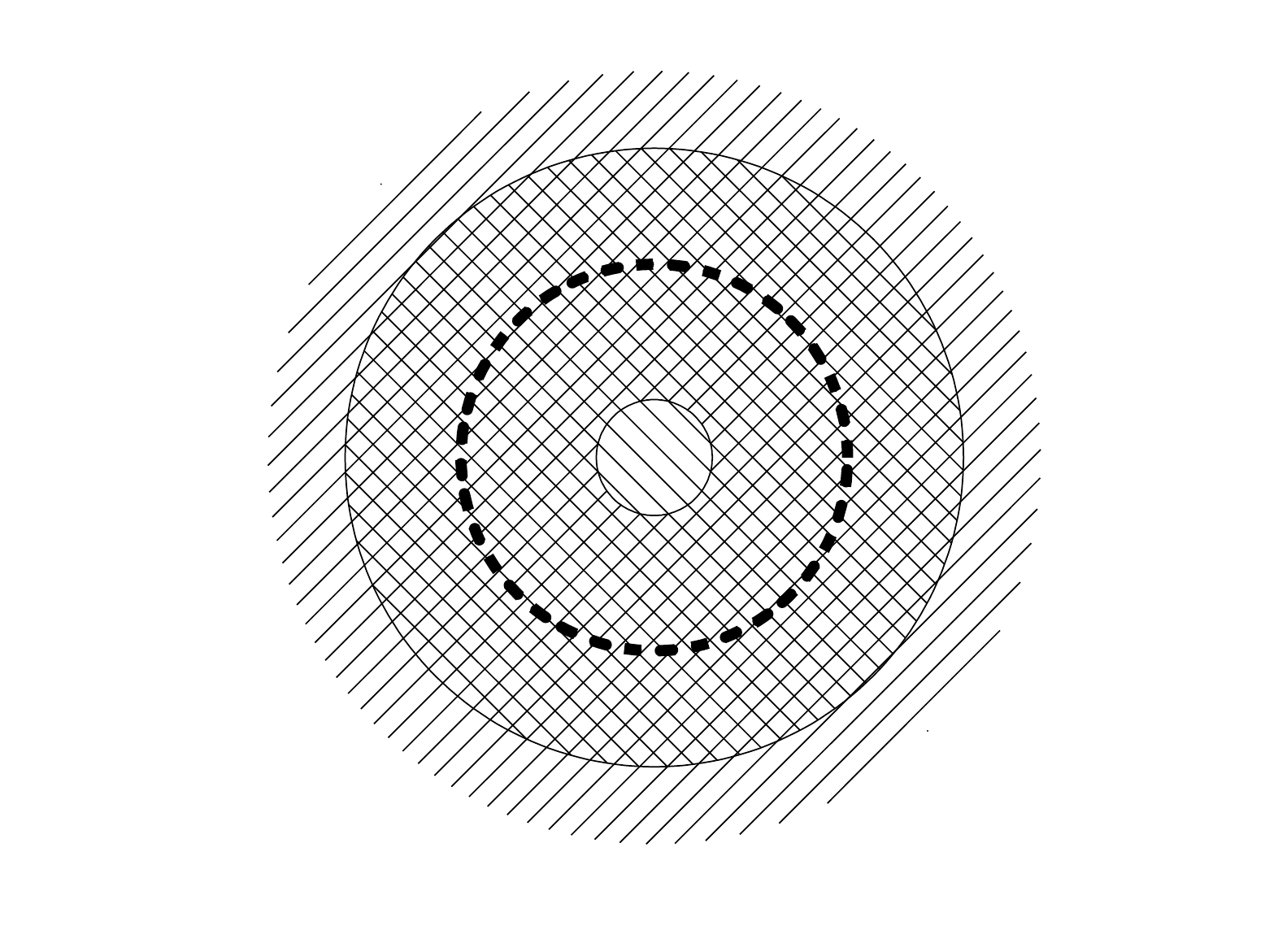}}}

\caption{ Effect of high and low values of $\frac{m_2}{m_1}$ for a fixed value
of $\frac{M_2}{M_1} \sim 1$. Left: high value of $\frac{m_2}{m_1}$. Right: low value
of $\frac{m_2}{m_1}$. The mixed region grows with decreasing $\frac{m_2}{m_1}$
and decreases with increasing $\frac{m_2}{m_1}$,  because of the enhanced
effect of mass segregation.}
\label{cartoon3}
\end{figure}

Firstly, let us consider models in which $\frac{m_2}{m_1} \gg 1$. In a region where both
components are present at comparable densities, there is a strong tendency
towards mass segregation. Therefore, in the region at which
$\frac{\rho_2}{\rho_1} \sim 1$, the ratio $\frac{\rho_2}{\rho_1}$  is a rapidly
decreasing function of the radius, i.e .the transition region is narrow. Inside this
region, $m_2$ dominates, and $m_1$ dominates outside. Clearly the radius at
which this region is located increases with $\frac{M_2}{M_1}$, and must be near
$r_h$ when $\frac{M_2}{M_1}=1$ (Fig. \ref{cartoon2}). Finally, for models in
which $\frac{m_2}{m_1} \not\gg  1$, the tendency towards mass segregation
decreases, the decrease of $\frac{\rho_2}{\rho_1} $ with $r$  is more gradual,
and the transition region is more extensive (Fig. \ref{cartoon3}). For the
same reason the regions dominated by a single component are more restricted than when
$\frac{m_2}{m_1} \gg 1$.

\subsubsection{Dependence on the number of heavy stars $N_2$ }\label{sec:DOH}
 The values of $N_2$ at $N_{crit}$ are given in Table \ref{table:N2}. The variation in $N_2$ is considerably less than that of $N_{crit}$. For large $\frac{m_2}{m_1}$ and fixed  $\frac{M_2}{M_1}$, the value of $N_2$ has approximately the same value, independent of $\frac{m_2}{m_1}$. Now, we give a possible interpretation of this empirical finding that the stability of the system is dominated by the heavy component. 
\begin{table}
\begin{center}
\caption{Number of heavy stars ($N_2$) at $N_{crit}$ in units of $10^4$ }
\begin{tabular}{ c  |c |c| c |c |c| c|c|c|}
\cline{2-9} 
& \bf{1.0}& 0.57    & 0.50 &   0.48  &  0.47 &   0.45    & 0.41   & 0.35  \\ \cline{2-9}
 & \bf{0.5} &   0.44 &   0.40 &    0.39  &  0.36 &   0.34 &   0.32 &   0.30 \\ \cline{2-9}
  $\frac{M_2}{M_1} $ & \bf{0.4}  &  0.38 &   0.38  &  0.35   & 0.34  &  0.36  &  0.29  &  0.26  \\ \cline{2-9}
   & \bf{0.3} & 0.34 &   0.33  &  0.32    & 0.31 &   0.29 &   0.27 &   0.25 \\ \cline{2-9}
    & \bf{0.2}& 0.27 &   0.28  &  0.26    &0.27  &  0.24  &  0.22  &  0.22  \\ \cline{2-9}
    & \bf{0.1}& 0.18 &   0.19  &  0.21 &    0.20 &   0.22 &   0.18 &   0.20 \\ \cline{2-9}
\cline{2-9}
                    &     & \bf{2}   & \bf{3}   & \bf{4}   & \bf{5}   & \bf{10}  & \bf{20}  & \bf{50} 	\\
\cline{2-9}
\label{table:N2}
\end{tabular}
\\
 $\frac{m_2}{m_1}$ 
\end{center}
\end{table}

Firstly, let us consider the case of $\frac{M_2}{M_1} \gtrsim 1$ and $\frac{m_2}{m_1}\gg1$ (Fig. \ref{cartoon2}, left). Within $r_h$ the heavy component dominates and most of the light component is removed to the outer halo. In this case, the light component acts as a container for the heavy component. Here, the stellar mass of the light component is not the most important factor, rather the most important factor is the overall mass of the container. If we were to replace this by an equal mass of stars with stellar mass $m_2$, the behaviour of the stars inside $r_h$ would be nearly the same, and so the value of $N_2$ at the stability boundary would be roughly the same as for a one-component model. Indeed, since part of the container consists of stars of mass $m_1$, this could also explain why the values of $N_2$ are in fact somewhat less than the value of $N_{crit}$ for a one-component system (i.e. 7000, \citet{Goodman1987}) and, in fact, why the critical value of $N_2$ is decreasing with decreasing $\frac{M_2}{M_1}$.  On the other hand, the fact that the critical value of $N_2$ is less than 7000 may also partly be due to the fact that the energy generation rate, equation (\ref{eq:energygenerationrate}), is smaller than that used by Goodman.  In his paper \citep[equation II.26]{Goodman1987} he shows implicitly that the critical value of $N$ is approximately proportional to the square root of the numerical coefficient in $\epsilon$.  At any rate, the arguments we have presented are  consistent with the results in the uppermost rows of Table \ref{table:N2}. 

Secondly, consider the case $\frac{M_2}{M_1} \lesssim 1$ (Fig. \ref{cartoon2}, right).  If the system is Spitzer unstable, the heavy component decouples from the light component and forms its own subsystem. This heavy subsystem can itself become gravothermally unstable and exhibit a temperature inversion in the same way as a one-component model.  In this case, however, there is not enough mass in the heavy component to dominate throughout the region within $r_h$, and so it is not quite as easy to relate this to the one-component case. Rather, we assume that the heavy component behaves like a detached one-component model. However, the basic conclusion is still the same, the stability of the model is determined by the heavy component. Since the heavy component is again sitting in the potential well of the lighter stars, it is easier for a nearly isothermal region to be set up in the heavier stars than if the entire system consisted of heavy stars, and we again expect $N_{crit}$ to correspond to a lower value of $N_2$ than in the one-component case. 

There is also a noticeable increase in the values of $N_2$ with decreasing $\frac{m_2}{m_1}$ in the top rows of Table \ref{table:N2}. There is currently no clear interpretation of this effect but it may possibly be related to the effect of mass segregation, as the region dominated  by the heavy component is larger for larger $\frac{m_2}{m_1}$ (see Fig.
 \ref{cartoon3}).

\subsubsection{Goodman's stability parameter}
\cite{Goodman1993} suggested that the quantity
\begin{center}
\begin{equation}\label{eq:eps1}
\quad
\qquad \qquad \qquad   \epsilon \equiv \frac{E_{tot}/t_{rh} }{E_c /t_{rc}}
\end{equation}
\end{center}
should indicate the stability universally, where $\log_{10}{\epsilon} \sim -2$ is the stability limit below which the cluster would become unstable. Here $E_{tot}$ is the total energy, $E_c$ is the energy of the core, $t_{rc}$ is the core relaxation time and $t_{rh}$ is the half mass relaxation time. \cite{KimLeeGood1998} carried out research using a Fokker-Planck model which seemed to support the condition, although the models they studied were all Spitzer stable. 

We have compared the values of $\epsilon$ found by \cite{KimLeeGood1998} to results obtained from the gas code (Table \ref{table:KLG}). All the models compared in Table \ref{table:KLG}, which are the same as those studied by \cite{KimLeeGood1998}, are stable in the post-collapse expansion as well as being Spitzer stable. An important difference between the Fokker-Planck model used by \cite{KimLeeGood1998} and the gas code used in this paper is that \cite{KimLeeGood1998} included an energy generation term in both components, whereas the gas code only contains an energy generation term in the heavier component. Therefore, it would be expected, in the case of the gas code, that the core would have to collapse further in order to generate the required amount of energy (from H\'{e}non's principle, see Section \ref{sec:DOI}). This could explain the differences in the values of $\frac{r_c}{r_h}$ in Table \ref{table:KLG}. However, as $\frac{M_2}{M_1}$ increases, the heavier component will dominate in the core and the energy generation of the lighter component will then become negligible. As can be seen in Table \ref{table:KLG}, there is good agreement between the two results for $\log_{10} \epsilon$ even though there are only small values of $\frac{M_2}{M_1}$. Also, it is possible that \cite{KimLeeGood1998} used a different definition of $t_{rc}$ than the one used in this paper (see equation \ref{eq:trc}). However, as there is such good agreement between the values in Table \ref{table:KLG}, it is unlikely that \cite{KimLeeGood1998} used a significantly different definition. Unlike the other models in the present paper, the models in Table \ref{table:KLG} are Spitzer stable. These runs have only been carried out in order to make a comparison of the calculation of $\epsilon$ and $\frac{r_c}{r_h}$ using the gas code with the results of \cite{KimLeeGood1998}. Next we will test the use of epsilon as a stability criterion for  Spitzer unstable cases.

\begin{table}
\begin{center}
 \caption{ Comparison of values of $\epsilon$ and $\frac{r_c}{r_h}$}
\begin{tabular}{ | c |  c | c | c | c |    }
    \hline
      $\frac{m_2}{m_1} $ & $\frac{M_2}{M_1} $ & $N$ & $Kim$ $et$ $al$ $\log{\epsilon}$ & $Gas$ $model$ $\log{\epsilon}$  \\ \hline
      2 & 0.02 & 3 $\times$ $10^{4}$ & -1.620 & -1.553 \\ 
      3 & 0.03 &3 $\times$  $10^4$              & -1.224 & -1.167 \\ 
      3 & 0.03 & $10^{5}$ & -1.597 & -1.544 \\
        \hline
\label{table:KLG}
\end{tabular}

\begin{tabular}{ | c |  c | c | c | c | }
    \hline
      $\frac{m_2}{m_1} $ & $\frac{M_2}{M_1} $ & $N$ & $Kim$ $et$ $al$ $\frac{r_c}{r_h}$ & $Gas$ $model$ $\frac{r_c}{r_h}$   \\ \hline
      2 & 0.02 & 3 $\times$ $10^{4}$ & 7.03 $\times$ $10^{-3}$& 4.86 $\times$ $10^{-3}$\\      3 & 0.03 &   3 $\times$ $10^4$              & 1.31 $\times$ $10^{-2}$& 0.91 $\times$ $10^{-2}$\\ 
      3 & 0.03 & $10^{5}$ & 5.38 $\times$ $10^{-3}$& 3.93 $\times$ $10^{-3}$\\
        \hline
\end{tabular}
\end{center}
\end{table}

We tested the stability criterion based on equation (\ref{eq:eps1}) for the subset of  models given in Table \ref{table:epss}. For each fixed $\frac{M_2}{M_1}$ and $\frac{m_2}{m_1}$, the values of $\epsilon$ were found to decrease with increased $N$ up until the post-collapse evolution became unstable. The values of $\log_{10} \epsilon$ given in Table \ref{table:epss} are the values for the run with the highest stable $N$. As can be seen from Table \ref{table:epss} the value of $\log_{10} \epsilon$ is indeed in the region of $ -2$. However, the limiting value of stable $\epsilon$ varies with $\frac{m_2}{m_1}$ and to a much lesser extent with $\frac{M_2}{M_1}$.

\begin{table}
\begin{center}
\caption{Value of $\log \epsilon$ for largest stable run}
\begin{tabular}{ c |c | c |c |c| c |}
                    & \bf{1.0} & -1.65  & -1.78  & -2.15    & -2.75   \\ \cline{2-6}
$\frac{M_2}{M_1}$   & \bf{0.5} & -1.68  & -1.90  & -2.15    & -2.75   \\ \cline{2-6}
                    & \bf{0.2} & -1.65  & -1.84  &  -1.95   & -2.48   \\ \cline{2-6}
                    &          & \bf{2} & \bf{5} & \bf{10}  & \bf{50} \\ \cline{2-6}
\label{table:epss}
\end{tabular}
\\ $\frac{m_2}{m_1}$ 
\end{center}
\end{table}

Now we shall try to improve upon the definition of $\epsilon$. In equation (\ref{eq:eps1}), $t_{rc}$ and $t_{rh}$ are defined by
\begin{equation}\label{eq:trc}
  t_{rc} =
\frac{0.34  \sigma^3_{c,2} }{G^2 m_2 \rho_{c,2}  \ln { \Lambda}}
\end{equation}
and
\begin{equation}\label{eq:trheq}
 t_{rh} =
\frac{0.138N^{\frac{1}{2}}r_h^{^{\frac{3}{2}}}}{(G\bar{m})^{\frac{1}{2}} \ln { \Lambda}}.
\end{equation}
Note that $t_{rc}$ was defined using the properties of the heavy component in the core rather than the averages of both components, as  the heavy component dominates in the core. However $t_{rh}$ (equation (\ref{eq:trheq})) depends on  $N$ and $\bar{m}$, which can vary dramatically with $\frac{m_2}{m_1}$ and $\frac{M_2}{M_1}$ for fixed $N_2$ whereas, as argued in Section \ref{sec:lsm}, the important criterion is the number of heavy stars. We suggest that a modified version of the Goodman stability parameter could be constructed using a relaxation time based on the heavy component in place of $t_{rh}$. For example, if $\frac{M_2}{M_2} \gtrsim 1$ and we assume that the heavier component dominates within $r_h$, then the properties of the system within $r_h$ would be roughly similar to that of a one-component system with the same total mass. We can attempt to treat the system as if it consisted entirely of the heavy component with an effective number of stars $N_{ef}=\frac{M}{m_2}$. The half mass relaxation time of this one-component system would be 

$$t_{rh,ef}=\frac{0.138N_{ef}^{\frac{1}{2}}r_h^{^{\frac{3}{2}}}}{(Gm_2)^{\frac{1
}{2}}\ln{\lambda N_{ef}}}=\Big(  \frac{1+\frac{M_2}{M_1}}{\frac{M_2}{M_1} +
\frac{m_2}{m_1}} \Big) \Big(  \frac{\ln{\lambda N}}{\ln{\lambda N_{ef}}} \Big) t_{rh}.$$ 

We can define a modified stability condition by replacing $t_{rh}$ with $t_{rh,ef}$ in the definition of $\epsilon$ which would then give the following condition $$\epsilon_2\equiv \frac{E_{tot}/t_{rh,ef} }{E_c /t_{rc}} .$$ The values of  $\log_{10}{\epsilon}$ and  $\log_{10}{\epsilon_2}$ are compared in Table \ref{table:tab3} for the case $\frac{M_2}{M_1}=1$. The values of  $\log_{10}{\epsilon_2}$ are in much better agreement with each other than those of $\log_{10} \epsilon$ and suggest that a better stability condition is $\log_{10} \epsilon_2 \approx -1.5$ rather than $\log_{10}\epsilon \simeq-2$. For the cases with $\frac{M_2}{M_1} \lesssim 1$ it is unclear how to define an appropriate relaxation time, and so we will not consider the modified stability condition for those cases.


\begin{table}
\begin{center}
\caption{   $\log{\epsilon}$ and $\log{\epsilon_2}$ for the case $\frac{M_2}{M_1}=1$}
\begin{tabular}{ c |c | c |c | c |} \hline
      $\log{\epsilon}$   & -1.65 & -1.78  & -2.15  & -2.75   \\ \hline
      $\log{\epsilon_2}$ & -1.51 & -1.39  & -1.53  & -1.56   \\ \hline
      $\frac{m_2}{m_1}$  & \bf{2}& \bf{5} & \bf{10}& \bf{50} \\ \hline
\label{table:tab3}
\end{tabular}
\end{center}
\end{table}


To summarise, the values of $\epsilon$ (and especially $\epsilon_2$) seem to give an indication of stability for the two-component models but the values of $\epsilon$ were found to change with different conditions (e.g $\frac{m_2}{m_1}$). The critical value of $\epsilon_2$ is much less variable.  The critical value of $\log_{10}{\epsilon}$ or $\log_{10}{\epsilon_2}$ is still to be tested for multi-component models, and this would be an interesting topic for further research.



\subsection{Weak oscillations}\label{sec:DOI}

\cite{Henon2} suggested that the energy generation rate of the core is determined by the requirement that it meets the energy demands of the rest of the cluster. This demand is normally thought of in terms of the energy flux at the half mass radius.  We shall refer to this as H\'{e}non's principle.  This principle, together with the notion of gravothermal instability, is the basis of the usual qualitative picture of gravothermal oscillations \citep{betsug1984}, which we now recap.

In a situation with very large $N$, the core has to collapse to a small size in order to meet the required energy generation. The steady state is gravothermally unstable, as there would be a large density contrast in a nearly isothermal region. If the core is generating more energy than can be conducted away, this would cause the core to expand, cool and reduce its rate of energy generation. If there is sufficient expansion, then the core would be cooler than its surroundings. This would result in the core starting to absorb heat. Since the core has a negative specific heat capacity, this would cause the core to expand further and became even cooler than before \citep{betsug1984}. Ultimately, however, the core must collapse again to meet the energy requirements of the rest of the cluster. Here, we adapt this explanation of gravothermal oscillations to the case of two-component clusters.  

\begin{figure}
\subfigure{\scalebox{0.45}{\includegraphics{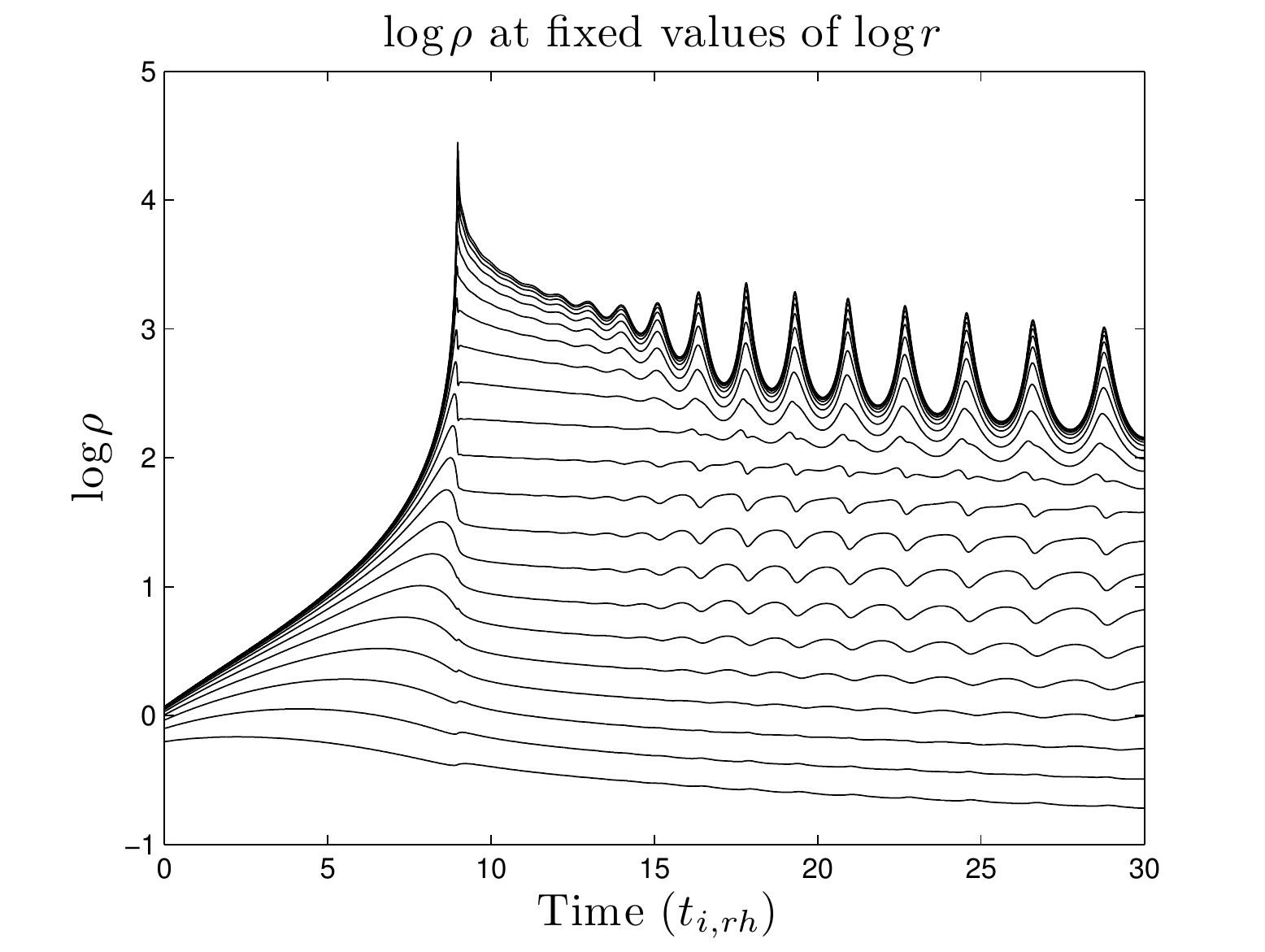}}}\quad
\subfigure{\scalebox{0.45}{\includegraphics{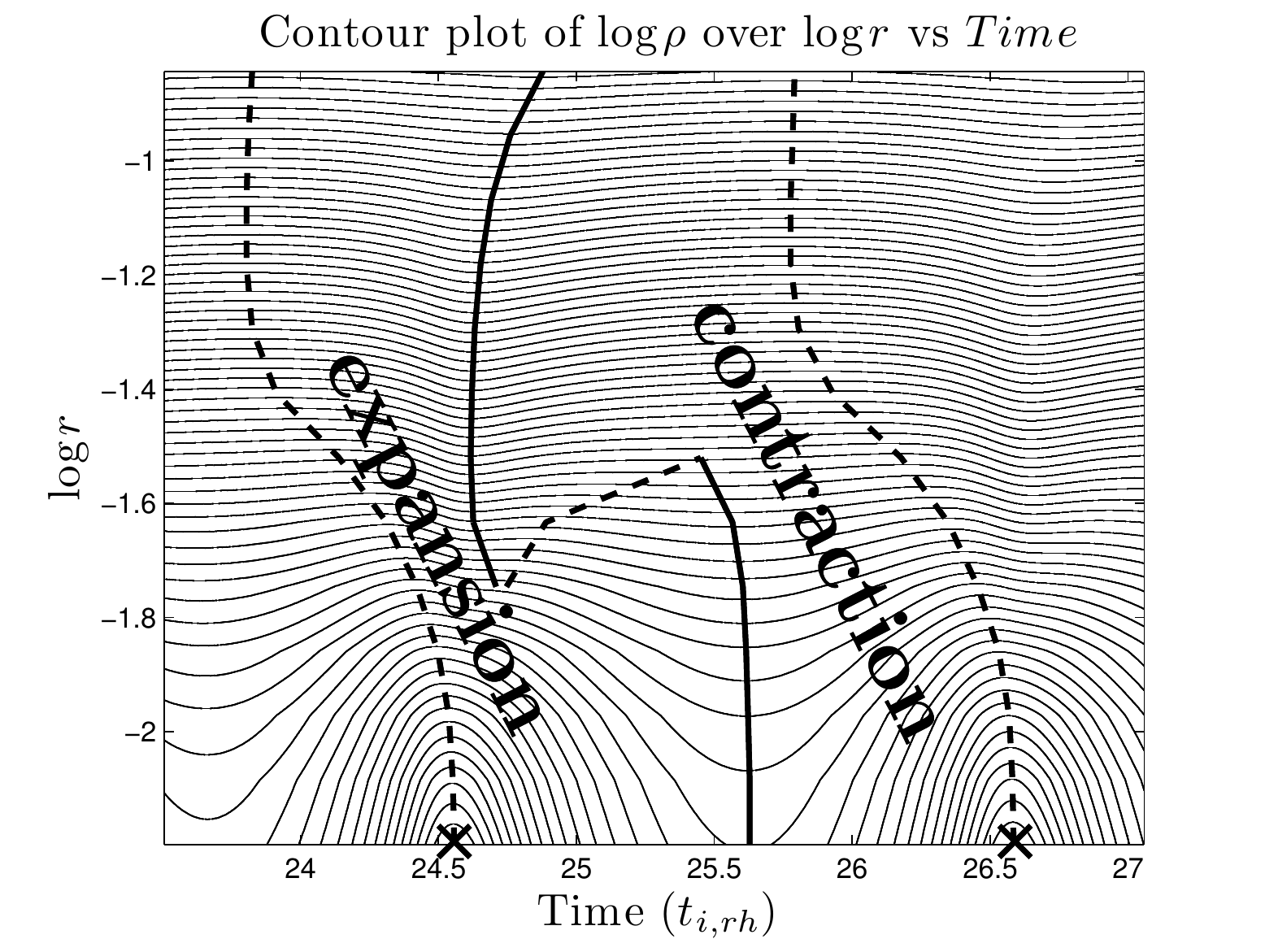}}}
\caption{Values of $\log{\rho}$ at fixed values of $\log{r}$ in the range  $-2.7621$ to $-0.5038$ in equal steps of size $0.1129$, for two-component models with $\frac{m_2}{m_1}=2$, $\frac{M_2}{M_1}=1$ and $N=2.0 \times 10^{4}$. Bottom: contour plot of $\log{\rho}$; the dashed lines represent the point of highest density reached locally over the time interval $23.5t_{i,rh}$ to $26.7t_{i,rh}$ and solid lines are the regions of lowest density reached between the dashed lines. $\times$ marks the points of core bounce, where the core stops contracting and starts expanding}

\label{fig:pc1}
\end{figure}

\begin{figure}
\centering
\subfigure{\scalebox{0.45}{\includegraphics{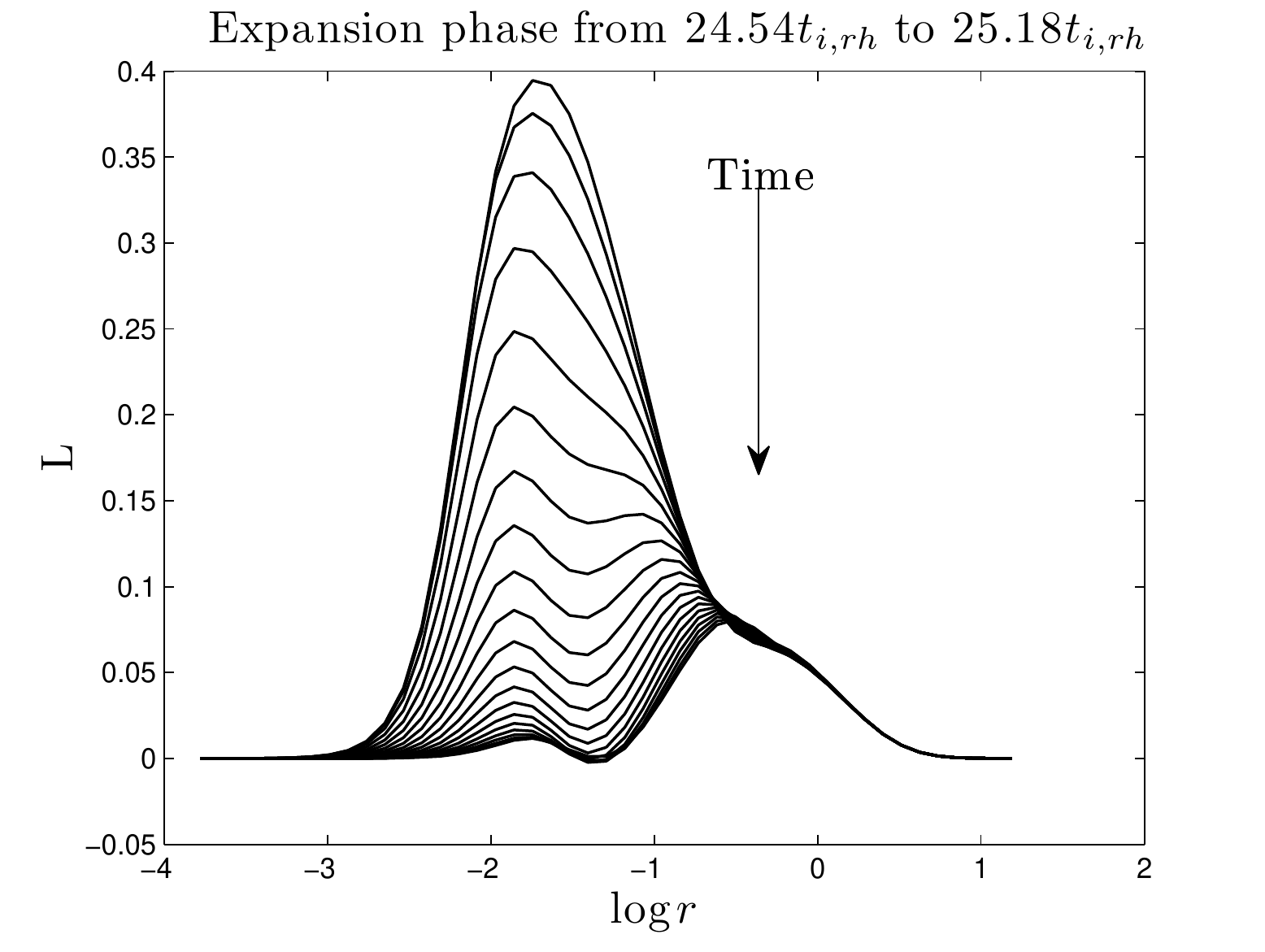}}}\quad
\subfigure{\scalebox{0.45}{\includegraphics{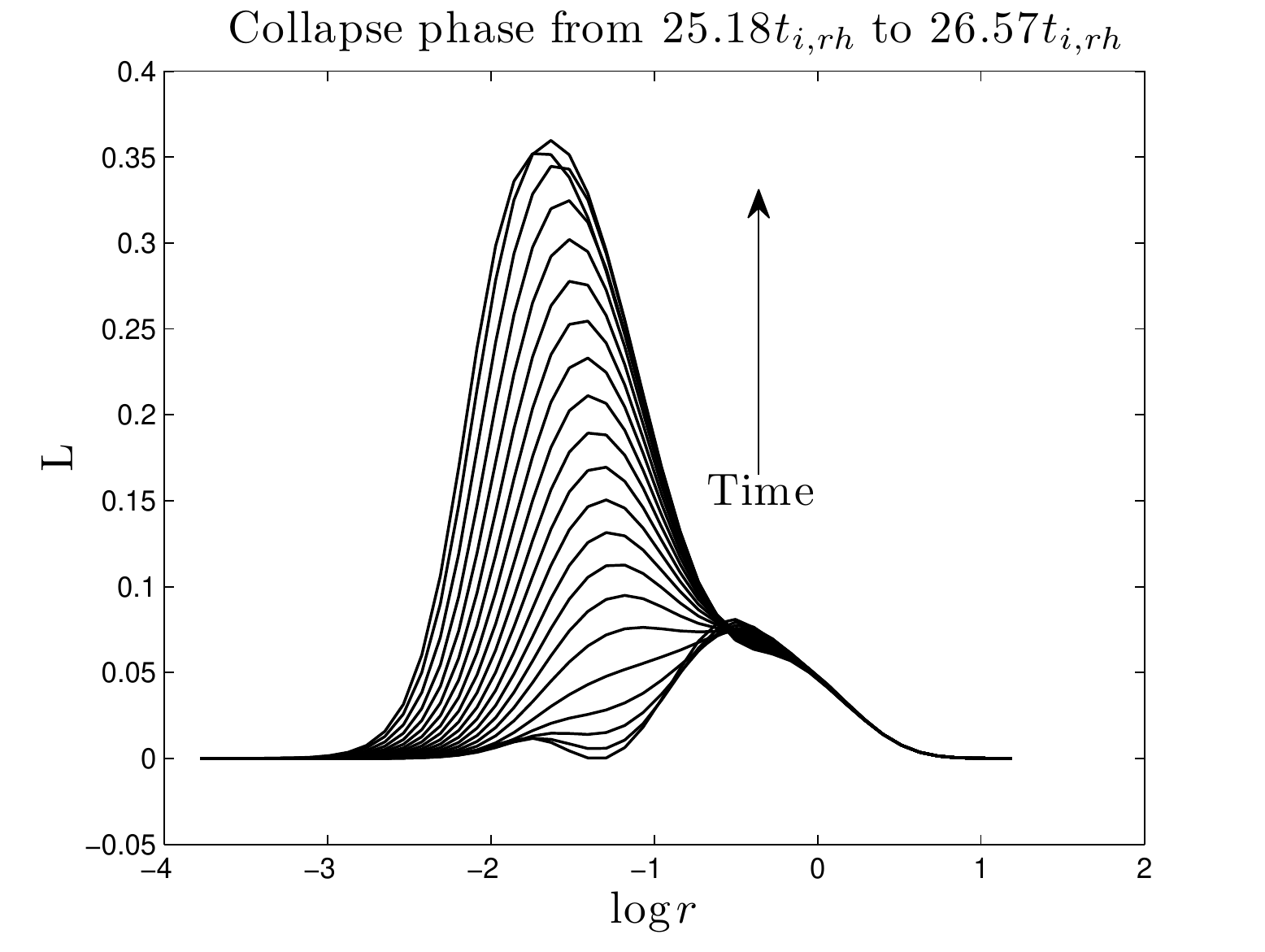}}}
\caption{Total energy flux in a two-component model with $\frac{m_2}{m_1}=2$, $\frac{M_2}{M_1}=1$ and $N=2.0 \times 10^{4}$.  The expansion phase from $24.54t_{i,rh}$ to $25.18t_{i,rh}$ is on the top and the contraction phase from $25.18t_{i,rh}$ to $26.57t_{i,rh}$ is on the bottom}
\label{flux}
\end{figure}

\begin{figure}
\subfigure{\scalebox{0.45}{\includegraphics{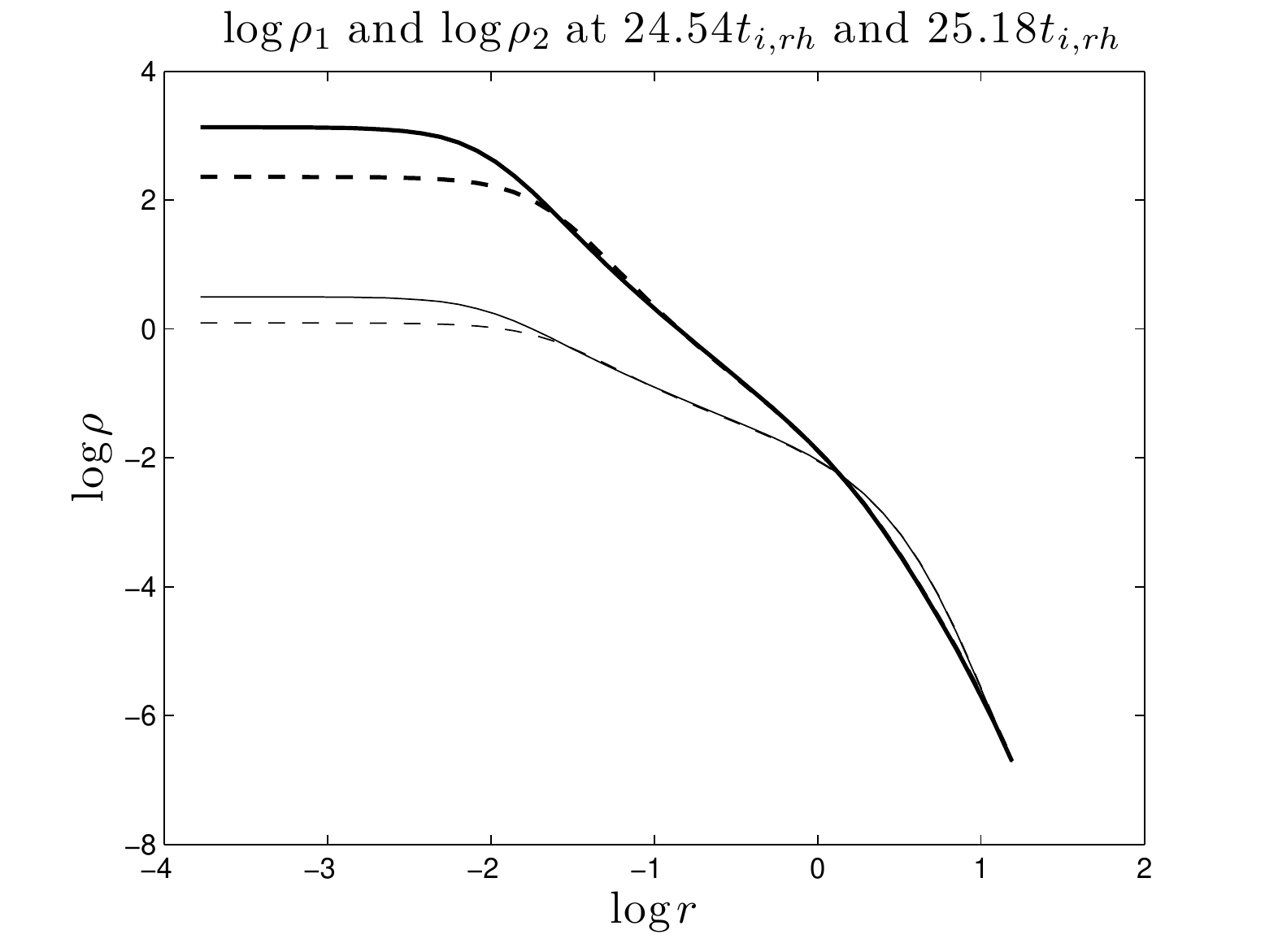}}}\quad
\subfigure{\scalebox{0.45}{\includegraphics{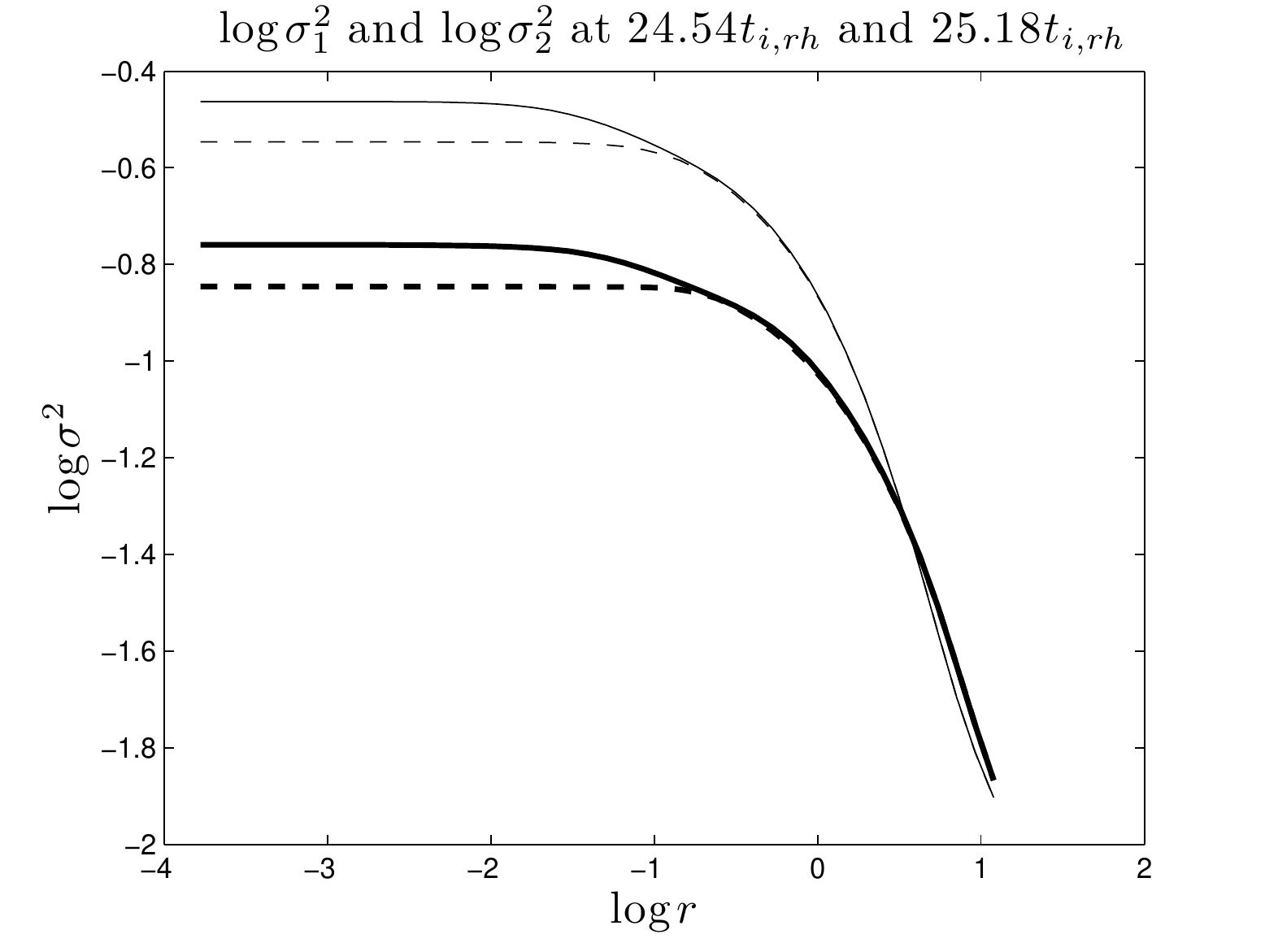}}}

\caption{Profiles of $\log_{10}(\rho)$ (top) and  $\log_{10}(\sigma^2)$ (bottom) for each component at maximum (dashed line) and minimum (solid line) expansion over times shown in Fig. \ref{flux}. The heavy (light) curves refer to the more (less) massive component}

\label{rho1rho2}
\end{figure}

In one-component gas models, as $N$ increases the instability first appears in the form of periodic oscillations\footnote{Strictly, only periodic if one scales out the steady expansion} \citep{Heggieramamani}. In order to study the instability for the case of weak or low amplitude oscillations in our two-component model, a model was chosen which demonstrated periodic oscillation with parameters $\frac{m_2}{m_1}=2$, $\frac{M_2}{M_1}=1$ and $N=2.0 \times 10^{4}$ (the value of $N_{crit}$ for $\frac{m_2}{m_1}=2$, $\frac{M_2}{M_1}=1$ is $1.7 \times 10^3$ from Table \ref{table:tab1}). Fig. \ref{fig:pc1} plots $\ln\rho$ at various fixed values of $\log r$ for this model. The total energy flux $L$ is shown in Fig. \ref{flux} over the particular expansion phase from $24.54t_{i,rh}$ to $25.18t_{i,rh}$ and the contraction phase from $25.18t_{i,rh}$ to $26.52t_{i,rh}$. Fig. \ref{rho1rho2} shows the profiles of  $\log{\rho}$ and $\log{\sigma^2}$   over the expansion phase from $24.54t_{i,rh}$ to $25.18t_{i,rh}$.

During the expansion of the core, the flux in the inner region (between $r_c$ and $r_h$) drops and eventually becomes negative (Fig. \ref{flux}, top) in a small range of the radius. At this point there is an inwards flux of energy to the core. Since the core has a negative heat capacity, it would be expected that this would enhance the negative flux and therefore the expansion. However, the expansion stops at this point. This is similar to behaviour observed by \cite{McMillan}. Now we explain why this happens. \cite{Henon2} argues that the flux at $r_h$ must be maintained, and  we note that there is always a positive flux at the half-mass radius $r_h$. Since the flux from the core becomes negative at some radius between $r_c$ and $r_h$, there must be a positive flux gradient in some region between the core and half mass radius. This can be seen in Fig. \ref{flux} (top) towards the end of the expansion and it continues into the early part of the contraction phase (Fig. \ref{flux}, bottom).

The flux gradient can be related to density via equation (\ref{eq:four}). As the heavier component dominates in the inner regions (see Fig. \ref{rho1rho2}), the main contribution to the flux is from the heavier component (i.e. $L \sim L_2$). Outside the core the energy generation will be negligible. Finally, the temporal change in $\ln{\rho}$ is greater than that in $\ln{\sigma^3}$. Taking all of this into account and rearranging equation (\ref{eq:four}) will result in the following: 
\begin{center}

\begin{equation}\label{eq:five} 
 \frac{1}{r^2 \rho_2  \sigma^2_2 }\frac{\partial L}{\partial r} \simeq  \Big( \frac{D}{Dt} \Big) \ln{\Big(\frac{\rho_2}{\sigma^3_2} \Big)} \simeq  \Big( \frac{D}{Dt} \Big) \ln{\Big({\rho_2}\Big)}.
\end{equation}
\end{center}

Since all of the coefficients of the flux gradient are positive the sign of flux gradient must be the same as that of the Lagrangian derivative of the density. Thus a positive radial flux gradient in space implies that the density is increasing with time. This can be seen in Fig. \ref{fig:pc1}, where the dashed lines mark the moment when the contraction becomes an expansion, and the solid lines mark the time when contraction resumes. It is clear that the contraction begins at large radii ($\log{r}\gtrsim -1.6$) while the core is expanding, and that this region of contraction propagates inwards at later times.  This can be related to the position of the positive gradient in Fig. \ref{flux} via the above equation (as long as the density is low enough that energy generation is negligible). Therefore, the collapse of the parts of the cluster between the core and $r_h$ starts while the core is expanding, and brings the expansion to a halt. Note that Fig. \ref{fig:pc1} is density plotted at fixed radius whereas time-derivatives in equation \ref{eq:five} are at fixed mass. Nevertheless in Fig. \ref{fig:pc1} we can also see that there are intermediate radii in which the density evolves in the opposite way from the core.

Although we have constructed the details of this description in the context of two-component models, nothing we have said depends entirely on this, and it seems likely that similar ideas will apply to one-component and multi-component models.

.


\section{ Core Collapse Time }\label{sec:coll}

While it may seem that the study of core collapse times is inappropriate in the context of gravothermal oscillations, it can be argued that the collapse phase of a gravothermal oscillation is not essentially different from the phenomenon of core collapse. Furthermore, another reason for its inclusion is that the evolution of isolated two-component models is an interesting research topic in its own right, and with the aim of constructing a comprehensive approximate theory of these models, studying the core collapse time is an appropriate first step.

The core collapse time $t_{cc}$ for a one-component cluster with Plummer model initial conditions has been found to be approximately $15.5t_{i,rh}$\citep{GaDy, HeggieHut2003} using various methods. \cite{Takahashi} found a longer $t_{cc}$ of $17.6t_{i,rh}$ with a one-component anisotropic Fokker-Planck code. However, the presence of a range of stellar masses can have a dramatic effect on the collapse time because of the process of mass segregation. The effect of mass segregation in multi-component models has been studied using Fokker-Planck calculations \citep{Murphyetal1990,ChreWein1990} and Monte Carlo methods \citep{Gurkanetal}. The effect of mass segregation in two-component models has already been studied extensively using direct \Nbody methods  \citep{khalisi}.

\begin{figure}
\includegraphics[scale=0.55]{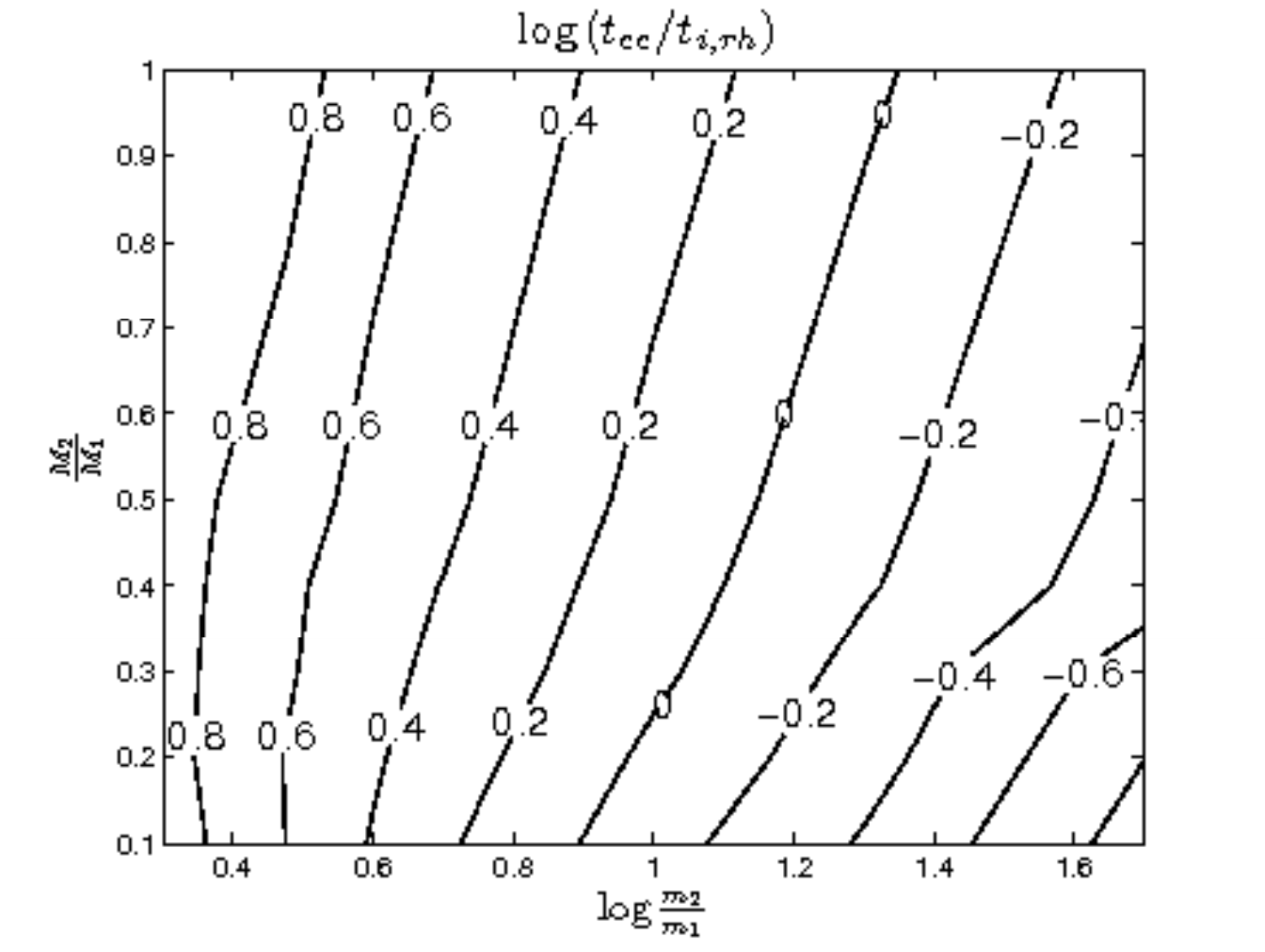}
\caption{Contours of $\log(\frac{t_{cc}}{t_{i,rh}})$ as a function of $M_2/M_1$ and $m_2/m_1$.}
\label{fig:logNcirt2}
\end{figure}

For the gas model runs discussed in Section 3, Table \ref{table:tab2} gives the values of the collapse time in units of the initial half mass relaxation time. Fig. \ref{fig:logNcirt2} shows a contour plot of log $t_{cc}$. The fastest collapse times occur with models of low $\frac{M_2}{M_1}$ and high $\frac{m_2}{m_1}$.   

%


\begin{table}
\begin{center}
 \caption{Collapse time $t_{cc}$ in units of the initial relaxation time} 
\begin{tabular}{ c  |c |c| c |c |c| c|c|c|}
\cline{2-9} 
                    & \bf{1.0} & 8.95 & 7.80 & 4.78 & 3.87  & 2.0   & 1.1  & 0.5  \\ \cline{2-9}
		    & \bf{0.5} & 7.80 & 4.78 & 3.45 & 2.75  & 1.38  & 0.72 & 0.35 \\ \cline{2-9}
 $\frac{M_2}{M_1} $ & \bf{0.4} & 7.58 & 4.43 & 2.89 & 2.49  & 1.23  &  0.66 &   0.31	\\ \cline{2-9}
                    & \bf{0.3} & 7.44 & 4.17 & 2.88 & 2.24  & 1.1   & 0.55   &  0.20	\\ \cline{2-9}
                    & \bf{0.2} & 7.42 & 3.89 & 2.65 & 1.97  & 0.91  &  0.47 & 0.16\\  \cline{2-9}
                    & \bf{0.1} & 8.1  & 3.95 & 2.4  & 1.7   &  0.75 & 0.38  & 0.13	\\  
\cline{2-9}
\cline{2-9}
                    &     & \bf{2}   & \bf{3}   & \bf{4}   & \bf{5}   & \bf{10}  & \bf{20}  & \bf{50} 	\\ \cline{2-9}
\label{table:tab2}
\end{tabular}
\\ $\frac{m_2}{m_1}$ 
\end{center}
\end{table}

\begin{figure}
\includegraphics[scale=0.55]{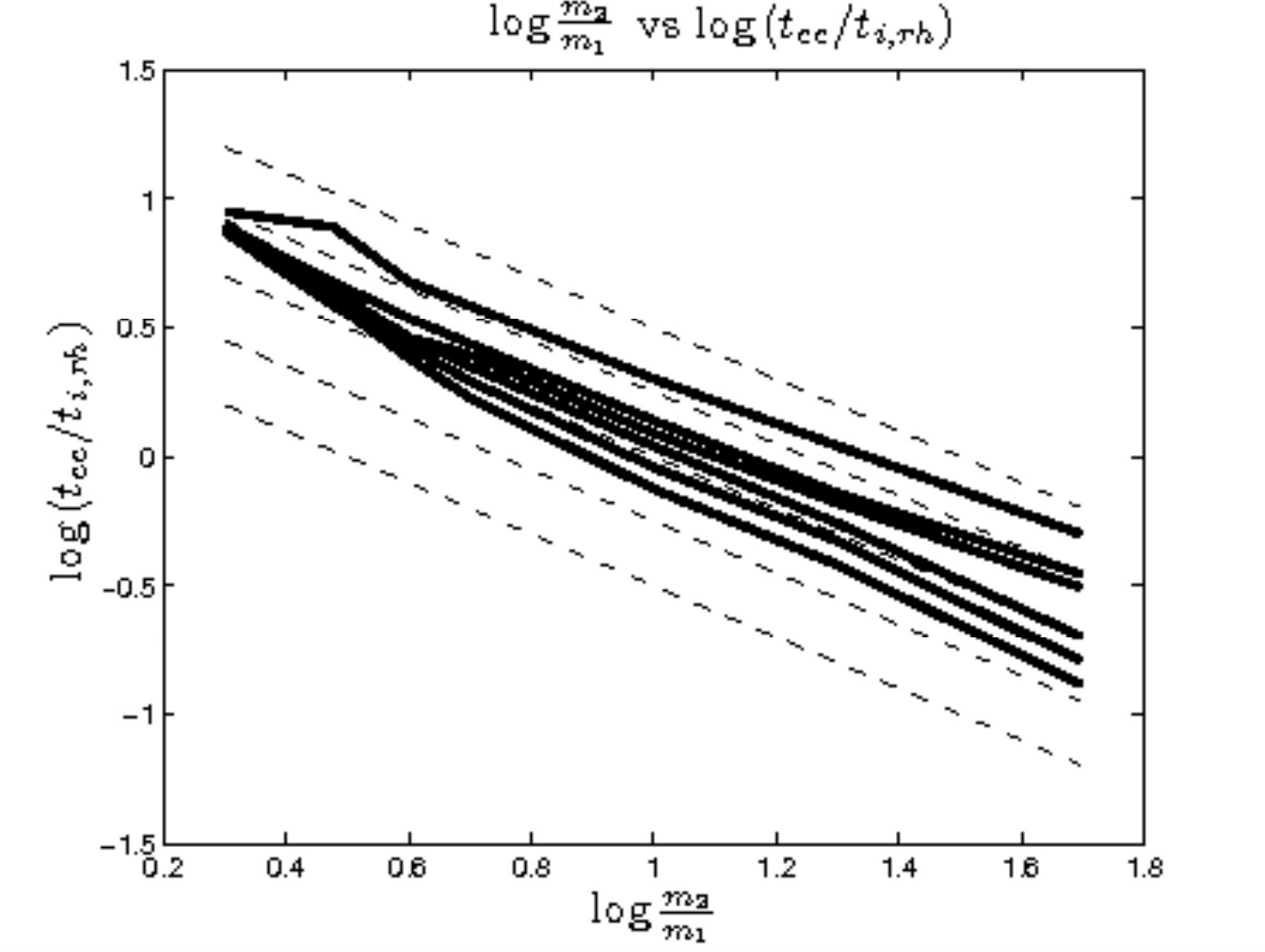}

\caption{ Solid lines are $\log(\frac{t_{cc}}{t_{i,rh}})$ vs $\log{\frac{m_2}{m_1}}$; from top to bottom
 $\frac{M_2}{M_1}= 1.0, 0.5, 0.4, 0.3, 0.2$ and $0.1$. Dashed lines are
$\log(k\frac{m_1}{m_2})$ vs  $\log{\frac{m_2}{m_1}}$  for various values of
$k$.  }
\label{logmu}
\end{figure}

For two-component systems, the timescale of mass segregation varies as $(\frac{m_2}{m_1})^{-1}$ \citep[and references therein]{Fregeau}. As mass segregation enhances the central density, it is expected that the mass segregation timescale is comparable with the timescale of core collapse.  Fig. \ref{logmu} compares the variation of the timescale of core collapse with the expected timescale of mass segregation. For the case of $\frac{M_2}{M_1} = 1.0$  (top line in Fig. \ref{logmu}) the collapse time indeed appears to vary as $(\frac{m_2}{m_1})^{-1}$. However, for lower values of $\frac{M_2}{M_1}$, the core collapse time decreases more quickly than for $(\frac{m_2}{m_1})^{-1}$. \cite{khalisi} also found a steeper decrease of the core collapse time in their study, for the case $\frac{M_2}{M_{i,tot}}=0.1$, where $M_{i,tot}$ is the initial total cluster mass. 

We can attempt to improve on these ideas at least qualitatively by considering in more detail a Spitzer unstable model. In that case, we can separate the pre-collapse evolution of the cluster into an initial mass segregation-dominated stage and a later gravothermal collapse-dominated stage, in which the centrally concentrated heavy component behaves almost as a one-component system thermally detached from the lighter component. We propose that this separation can be located via the minimum of the rate of change of central density ratio (i.e $\min\big\{{\frac{d}{dt}\big( \frac{\rho_2}{\rho_1}}\big)\big\}$). The reasoning behind  this is as follows: as time passes, the increase in the density ratio caused by mass segregation starts to slow due to a combination of decreasing relative density and increasing temperature of the lighter component in the central regions. We assume that it is at this point that the gravothermal collapse of the heavier component becomes the dominant behaviour of the system. The gravothermal collapse in the heavy component increases the temperature of the heavy component, and because the light component absorbs energy from the heavy component, the collapse of the heavy component causes a deceleration in the collapse of the light component. This in turn enhances the rate of increase in the density ratio. 

\begin{figure}
\subfigure{\scalebox{0.35}{\includegraphics[angle=270]{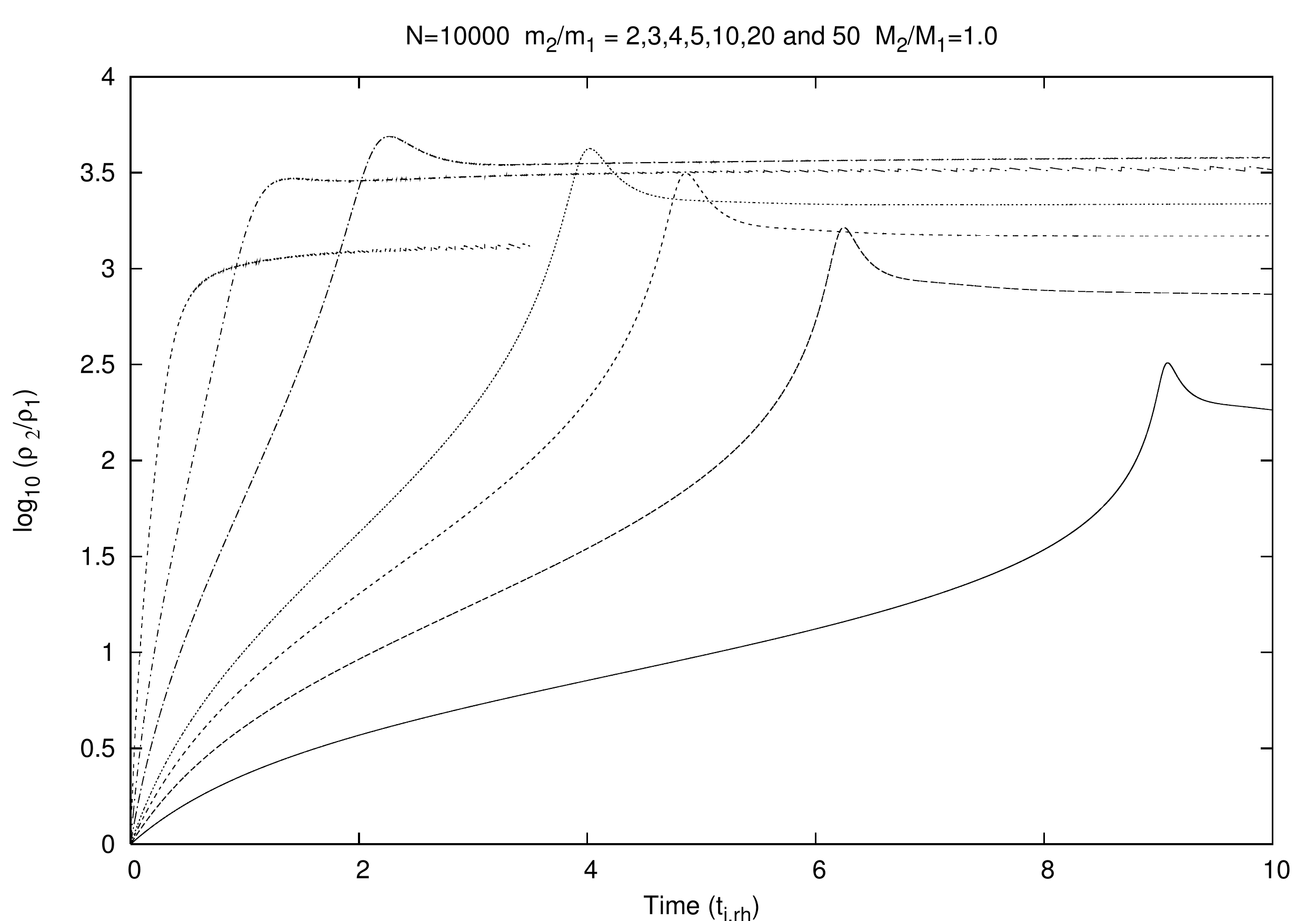}}}\quad
\subfigure{\scalebox{0.35}{\includegraphics[angle=270]{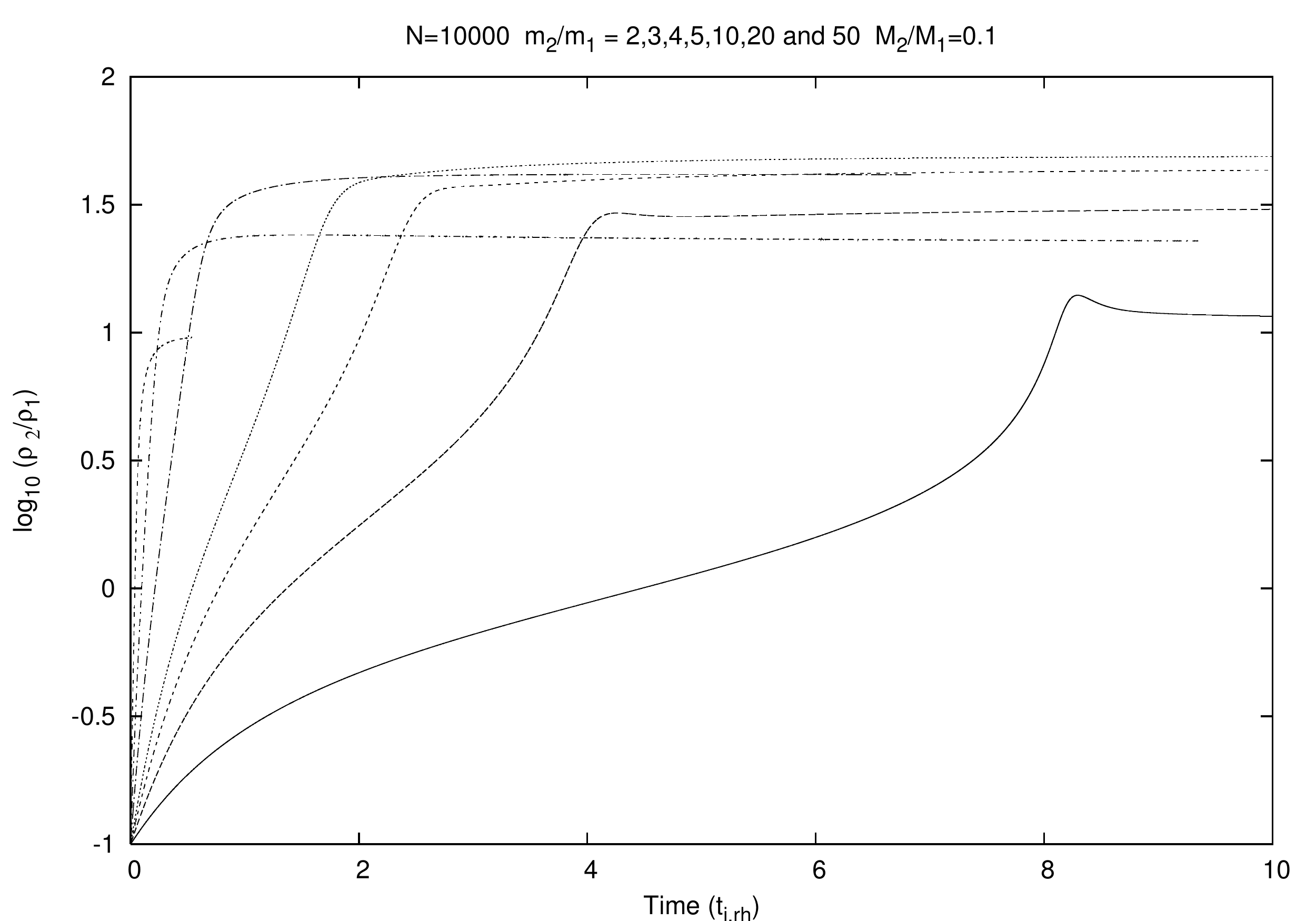}}}
\caption{$\log_{10}(\frac{\rho_2}{\rho_1})$ vs time (in units of $t_{i,rh}$) for the case of
$\frac{M_2}{M_1}=1$ (top) and  $\frac{M_2}{M_1}=0.1$ (bottom) Curves from
bottom to top are $\frac{m_2}{m_1}= 2, 3, 4, 5, 10, 20$ and $50$ }
\label{rhooverrho}
\end{figure}

Fig. \ref{rhooverrho} shows the density ratio ${\rho_2}/{\rho_1}$ vs
time for $N=10000$ and $\frac{M_2}{M_1}=1,0.1$.   For the case of
$\frac{m_2}{m_1}=2$ (the lowest curve) there is a clear distinction
between the part before the point of inflection at about
$\frac{t}{t_{i,rh}} = 5$ (i.e the initial mass segregation phase) and the part after the point of inflection (i.e the gravothermal collapse phase).  As $\frac{m_2}{m_1}$ increases the initial phase dominated by mass segregation becomes more substantial and eventually the initial mass segregation phase brings the system all the way to core bounce. However, as $N$ increases, binary energy generation becomes less efficient relative to the energy demands of the cluster \citep{Goodman1987}. Therefore, the core needs to reach a higher density at core bounce for larger $N$. As the initial phase of mass segregation is self limiting for the reason given above, mass segregation cannot increase the central density beyond a certain point. Therefore, it would be expected that the gravothermal collapse dominated phase must eventually return with increasing $N$ for any given $\frac{M_2}{M_1}$ and $\frac{m_2}{m_1}$.   

%



\section{Direct $N$-body}

\cite{betsug1985} compared $N$-body systems to gaseous models using a
direct $N = 1000$ model. Even though the value of $N$ is small by
today's standards there was still fair agreement during the
pre-collapse phase. There were large statistical fluctuations in the
post-collapse phase and this was most likely due to the small particle
number. However, it is still important to confirm a sample of the
results of the gas model by using a direct $N$-body code.

 The case of $\frac{m_2}{m_1}=2$ $\frac{M_2}{M_1}=1$ was chosen
 because it had the smallest value of $N_{crit}$. The values of $N$
 used for these runs were $8k,16k,32k$ and $64k$. The collapse times
 of the runs in \Nbody  units  \citep[see][]{Heggiemathieu1986} and
 units of $t_{i,rh}$ are given in Table \ref{table:tabdn}. The average
 collapse time measured in units of $t_{i,rh}$ is about $7.5$ which is
 lower than the predicted value of $8.95$ in Table \ref{table:tab2}. 
The difference in collapse time could be because of the approximate
 treatment of two-body relaxation in the gas model,  the neglect of
 escape, or parameter choices in the gas code (Section \ref{sec:gasmodel}).


\begin{table}
\begin{center}
 \caption{Collapse time $t_{cc}$ } 
\begin{tabular}{ | c | c | c | c| c }
\hline
$N$  &$8k$ &$16k$ & $32k$ & $64k$  \\  \hline
 \Nbody units &  $1160$     &  $1990$     &  $3480$   &  $  6380  $    \\ \hline
$t_{i,rh}$ & $7.76$ & $7.56$ & $7.41$ &  $7.51$ \\ \hline
\label{table:tabdn}
\end{tabular}

\end{center}
\end{table}

\begin{figure}
\includegraphics[scale=0.55]{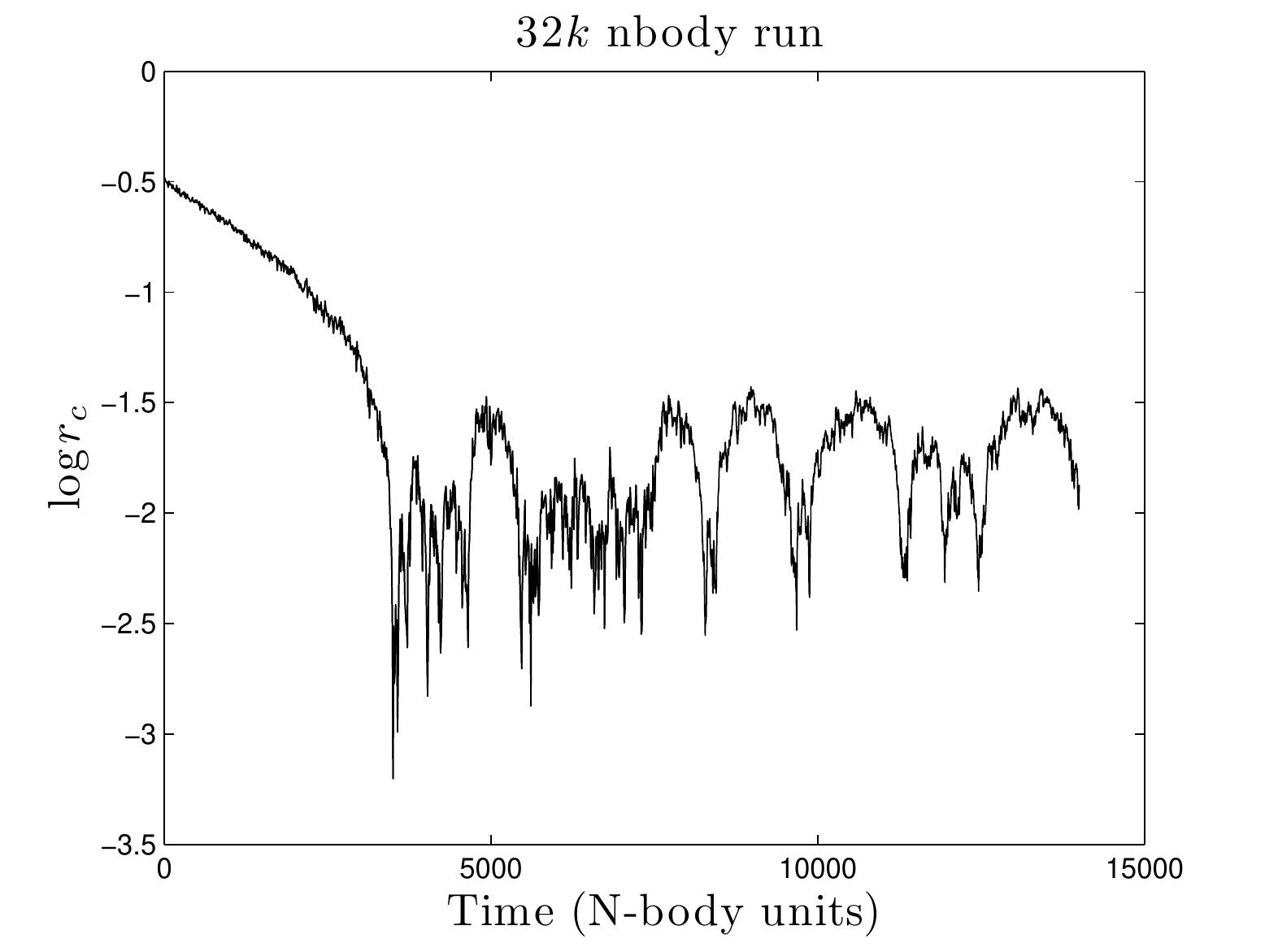}
\caption{$\log_{10} r_c$ vs time, where  $r_c$ is defined as in NBODY6, $\frac{m_2}{m_1}=2$, $\frac{M_2}{M_1}=1$ and $N$ $=$ $32k$. }
\label{nbody32}
\end{figure}
\begin{figure}

\subfigure{\scalebox{0.55}{\includegraphics{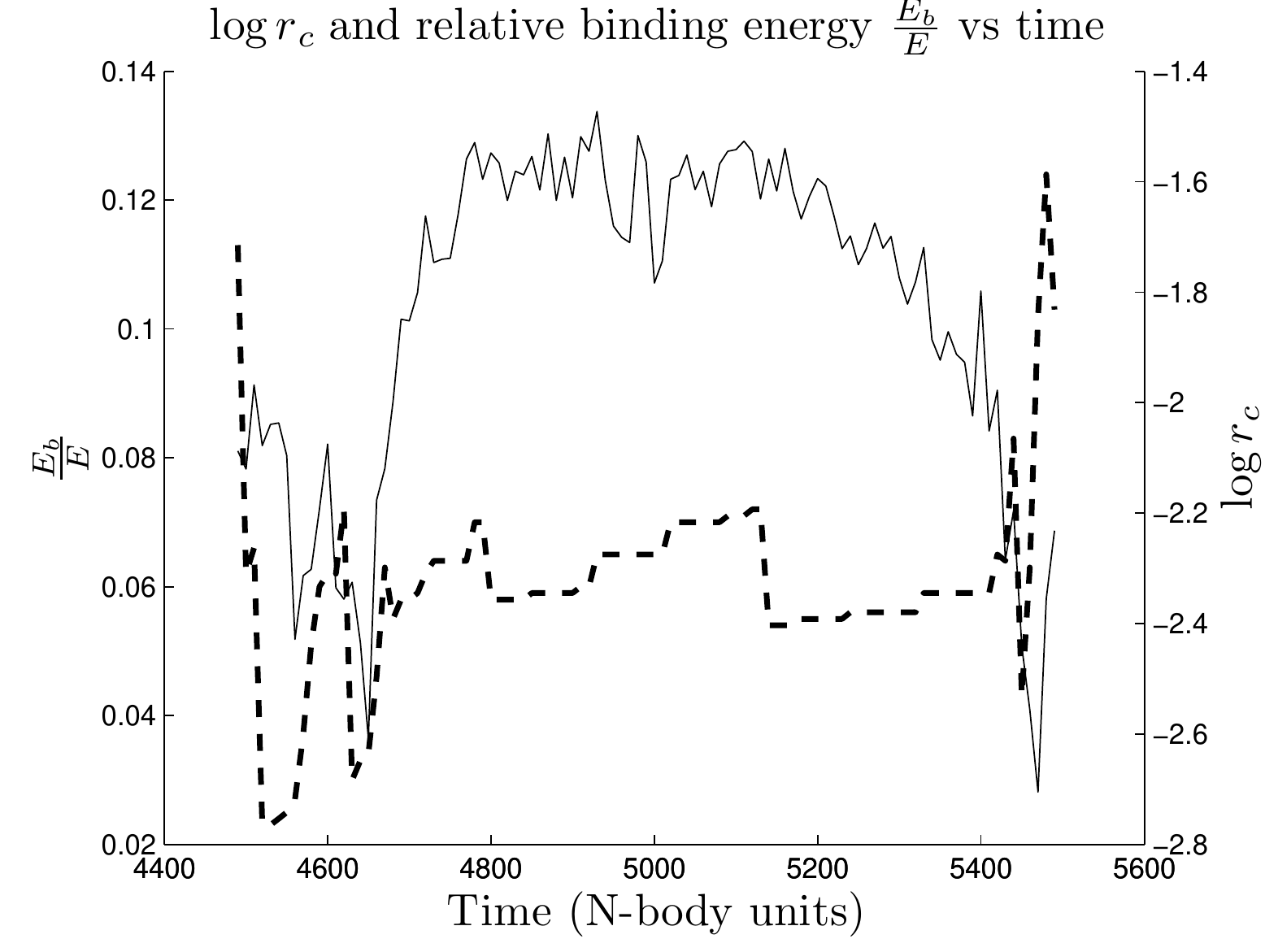}}}\quad
\subfigure{\scalebox{0.55}{\includegraphics{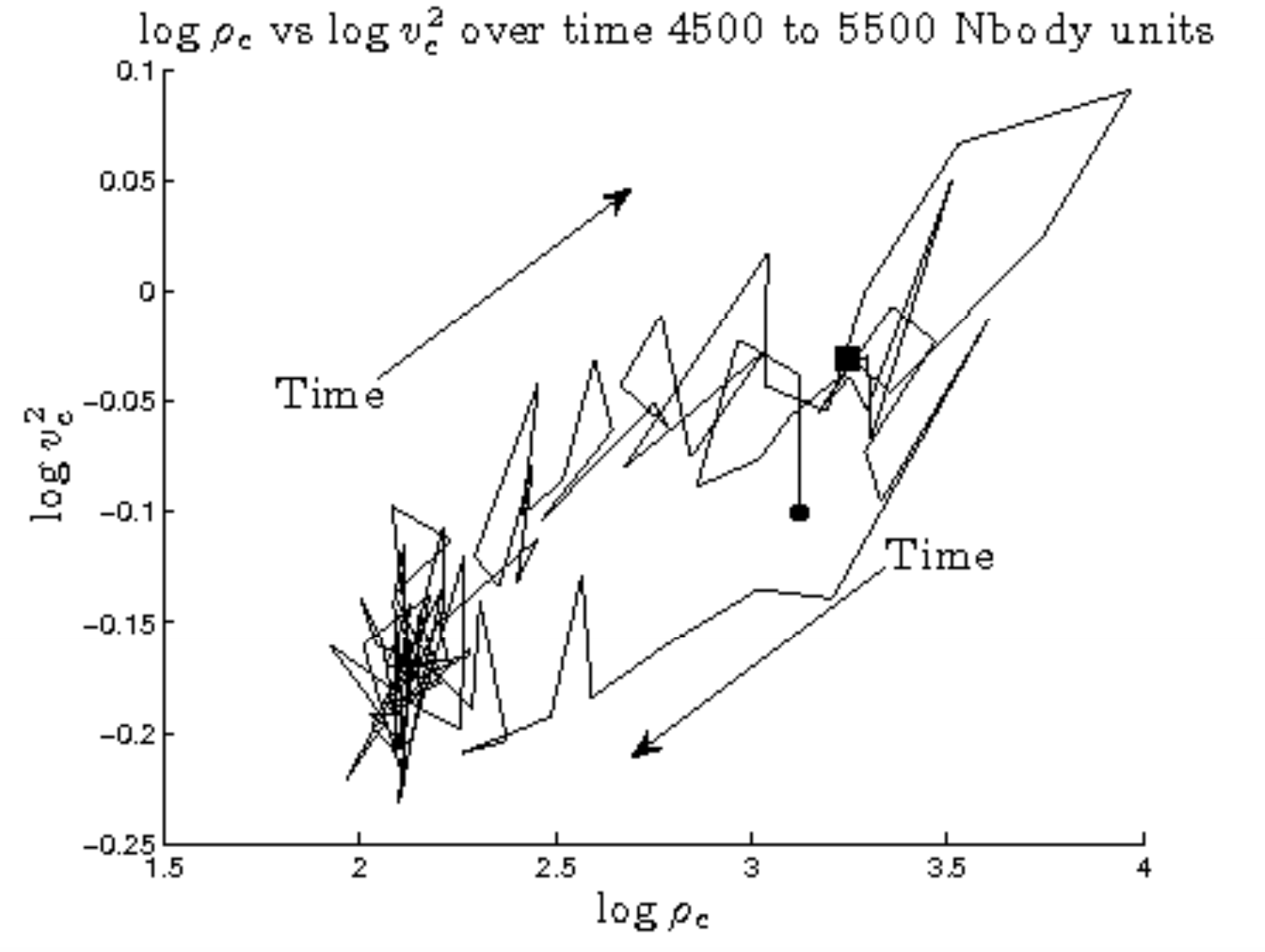}}}

\caption{$32k$ $N$-body results. Top: relative binding energy $\frac{E_b}{E}$ compared to $\log{r_c}$ over time 4500 to 5500 \Nbody units. Bottom: $\log{r_c}$ vs $\log{v_c^2}$ over the same time period, where, $\bullet$ and $\blacksquare$ represent the starting and finishing points.}
\label{nbody32gto} 
\end{figure}

\begin{figure}
\subfigure{\scalebox{0.55}{\includegraphics{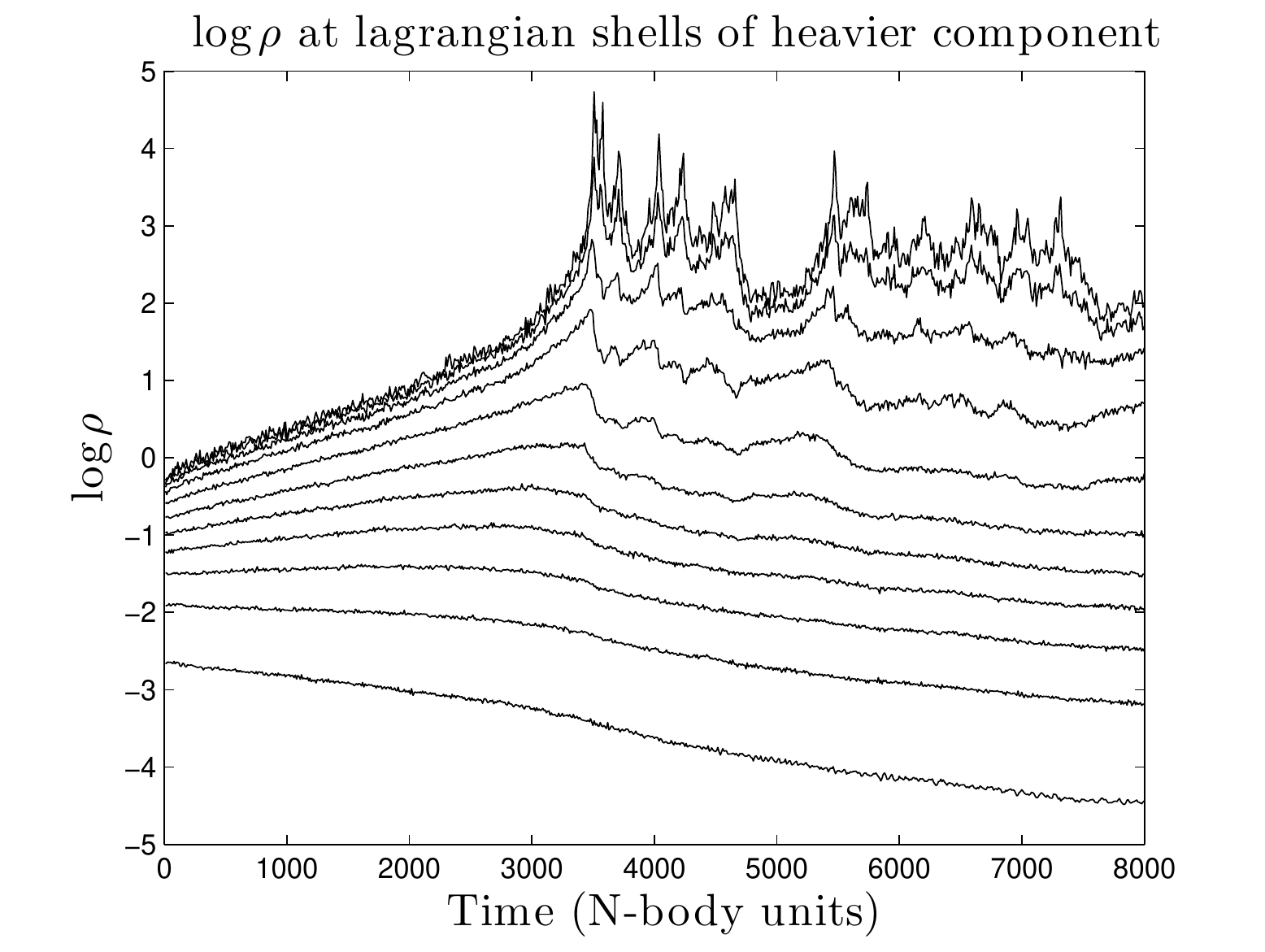}}}
\caption{$32k$ $N$-body results. $\log{\rho}$ in Lagrangian shells of 1, 2, 5, 10, 20, 30, 40,
  50, 62.5, 75 and 90 percent mass in the heavier component. It can be
  seen that the collapse at $5400$ starts further out (while the core
  is still expanding) and propagates towards the core.}
\label{nbody32lagr}
\end{figure}

For the case of the runs with $N$ equal to $8k$ and $16k$ no behaviour
was found which could be described as gravothermal oscillation.  This
is in agreement with the gas code, which gave  $N_{crit} =
17000$. However, the $32k$ case does show a cycle of expansion and contraction of the
core over the time interval $4500$ to $5500$ \Nbody units (see
Fig. \ref{nbody32}). In order to check that the expansion was not
driven by sustained binary energy generation, we consider the
evolution of the relative binding energy $\frac{E_b}{E}$, where $E_b$
is the total binding energy of the binaries and $E$ is the absolute
value of the total energy of the cluster, over this time period. This
is plotted in Fig. \ref{nbody32gto} along with the core radius. There
are small changes in the binding energy of binaries over this period,
decreases as well as increases, but this cannot fully account for the
expansion phase that is observed, as there are other periods with
similar binary activity in which no sustained expansion occurs. Also,
the time scale of the expansion is much longer than the relaxation
time in the core ($\sim$ 0.5 in \Nbody time units). Therefore, we
assume that the expansion must be driven by phenomena outside the
core, and gravothermal behaviour is a plausible explanation.  

\begin{figure}
\subfigure{\scalebox{0.55}{\includegraphics{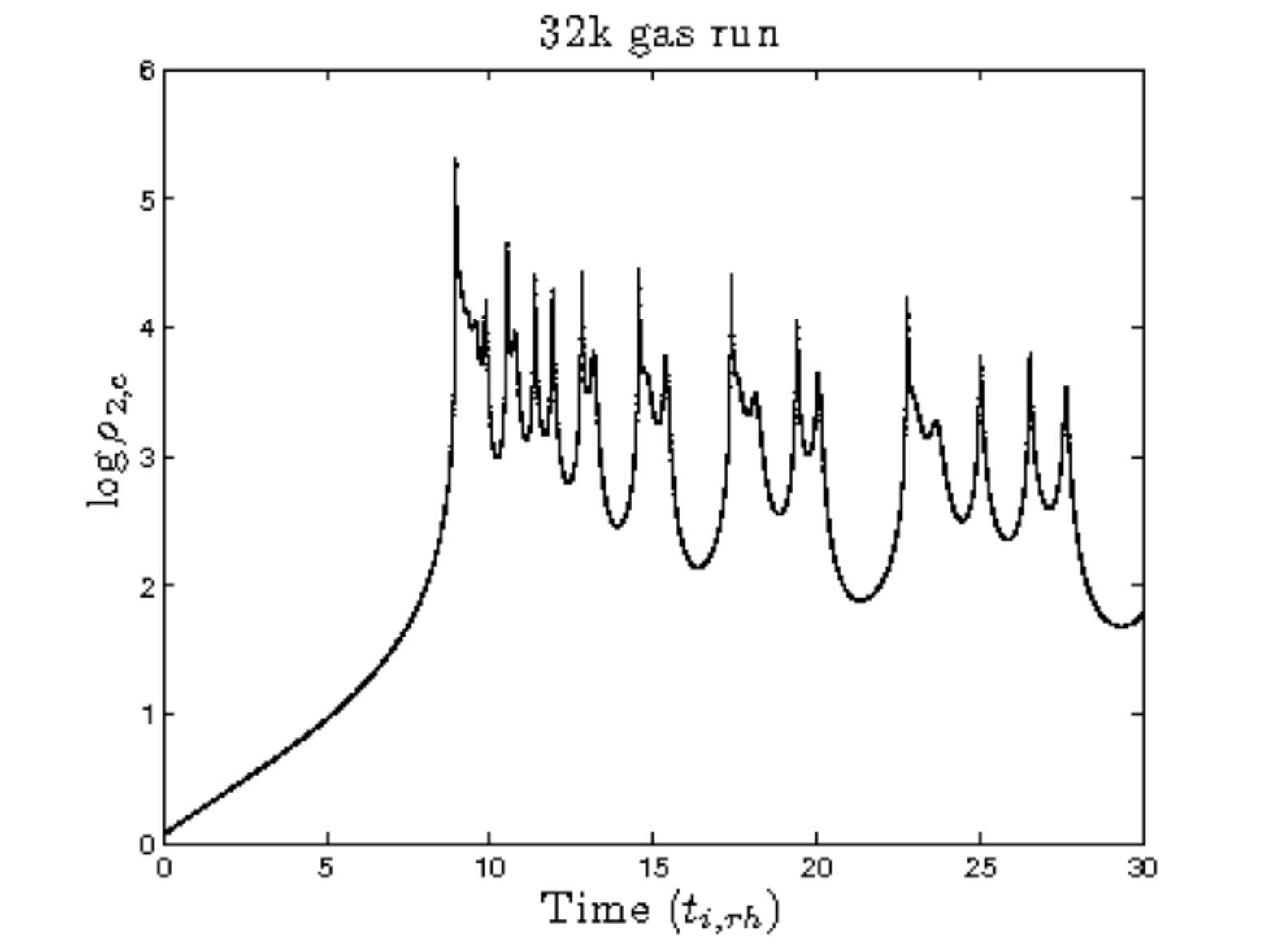}}}\quad
\subfigure{\scalebox{0.55}{\includegraphics{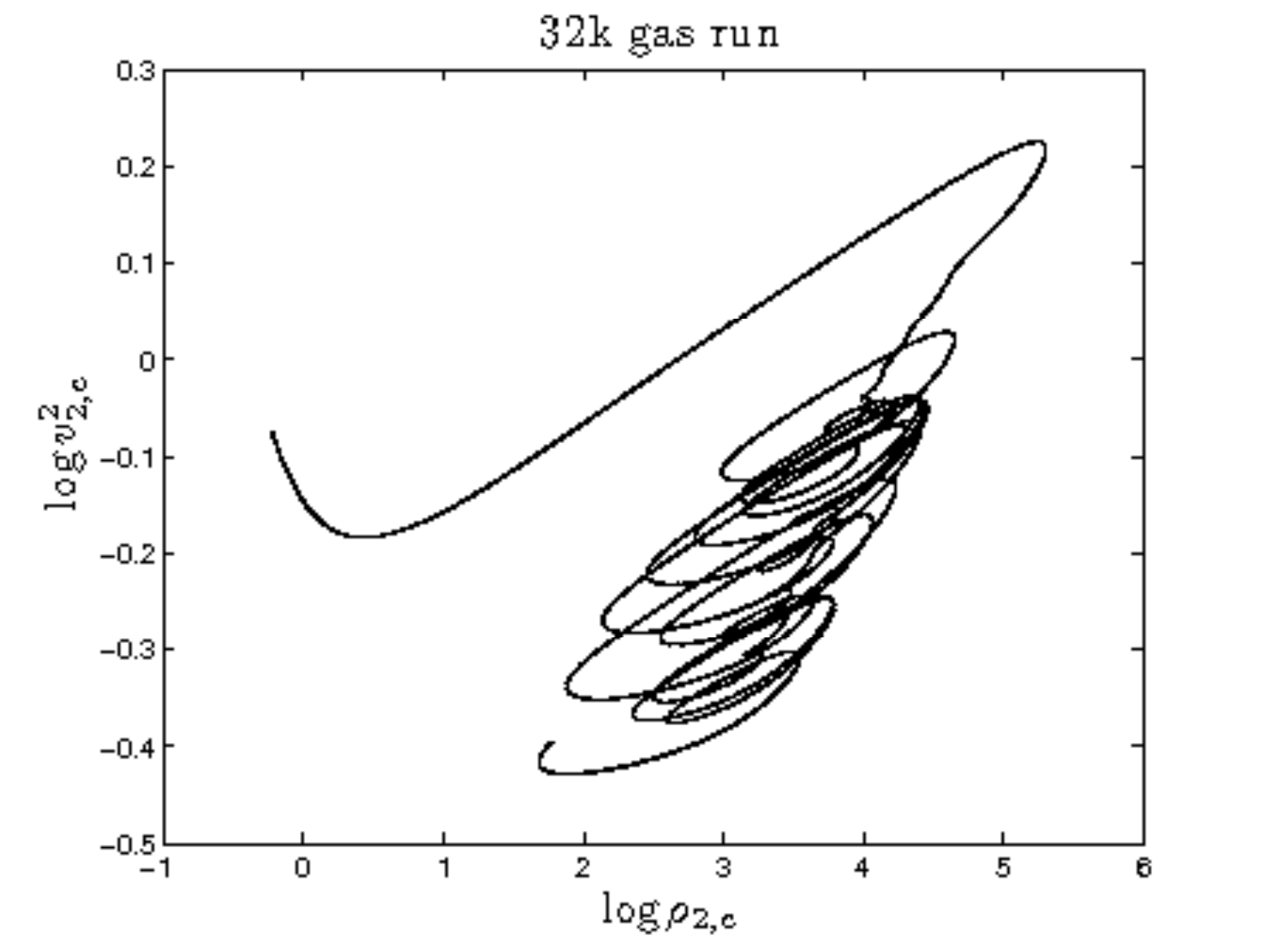}}}
\caption{Evolution of the central properties of the heavy component in a 32k gas model. Top: $\log{\rho_{2,c}}$ vs time $($units $t_{i,rh})$, bottom: $\log{\rho_{2,c}}$ vs $\log{v_{2,c}^2}$.  All cycles are clockwise.  The initial drop in $\log{v_{2,c}^2}$ results from the two components trying to achieve thermal equilibrium}
\label{gas32gto}
\end{figure}

Several other pieces of evidence point to this
conclusion. Fig. \ref{nbody32lagr} shows the density in Lagrangian
shells of the heavier component.  As discussed in Section
\ref{sec:DOI} (e.g. Fig. \ref{fig:pc1}, top) the region further away
from the core is seen to contract while the core expands.  Also, in
the cycle of $\ln{\rho_c}$ vs the core velocity dispersion
$\log{v_c^2}$,  the temperature is lower during the expansion where
heat is absorbed and higher during the collapse where heat is released
(Fig. \ref{nbody32gto}, bottom). This is similar to the cycles found by \cite{Makino1996} for one-component models and is another sign of gravothermal behaviour. The results from the $32k$ gas run are shown in Fig. \ref{gas32gto} for comparison.

\begin{figure}

\subfigure{\scalebox{0.55}{\includegraphics{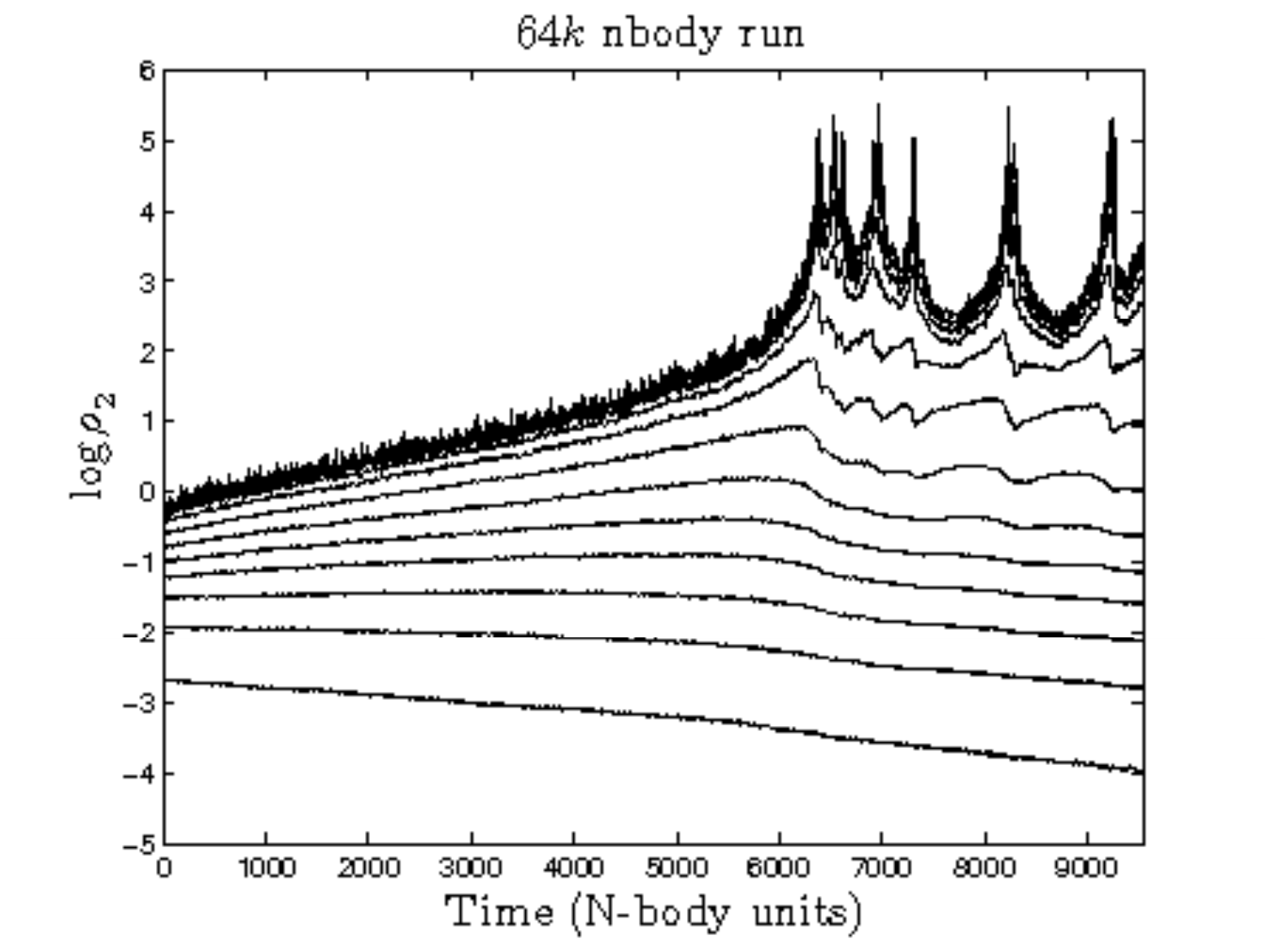}}}\quad
\subfigure{\scalebox{0.55}{\includegraphics{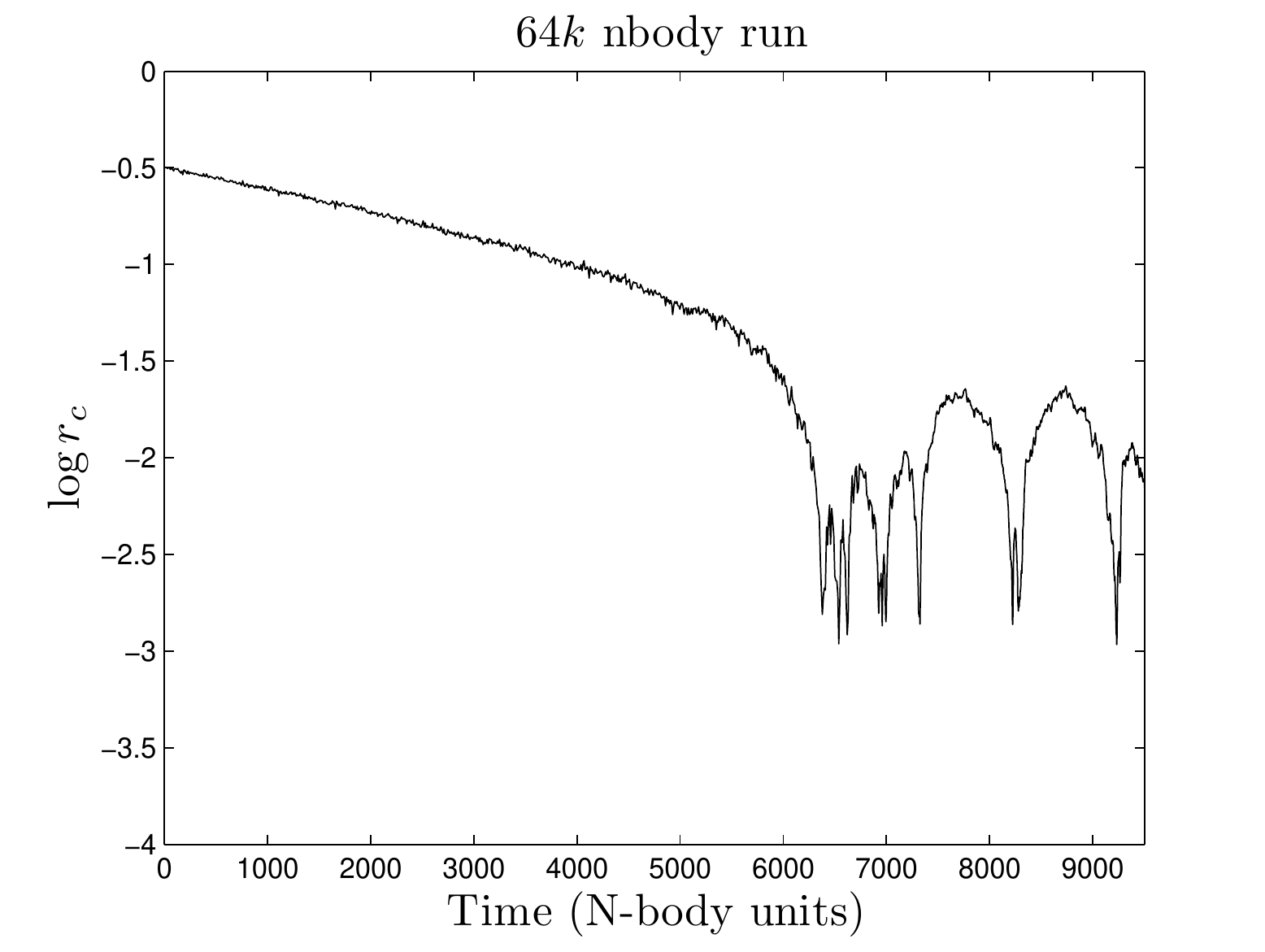}}}\quad

\caption{64k \nbody run. Top: $\log{\rho_2}$ at various Lagrangian shells in the heavier component, bottom: $\log{r_c}$ vs time}
\label{fig:64k}
\end{figure} 

\begin{figure}

\subfigure{\scalebox{0.55}{\includegraphics{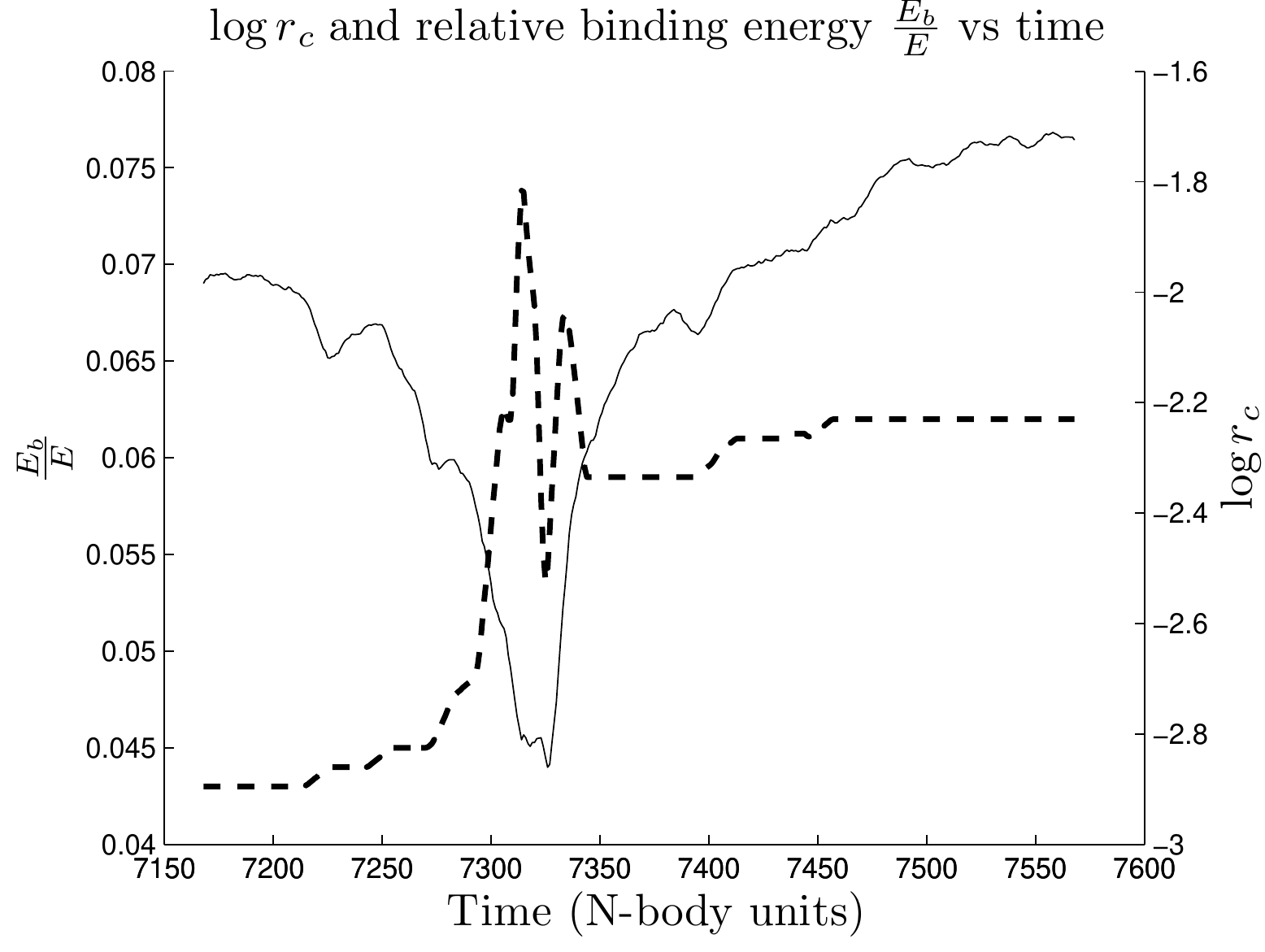}}}\quad
\subfigure{\scalebox{0.55}{\includegraphics{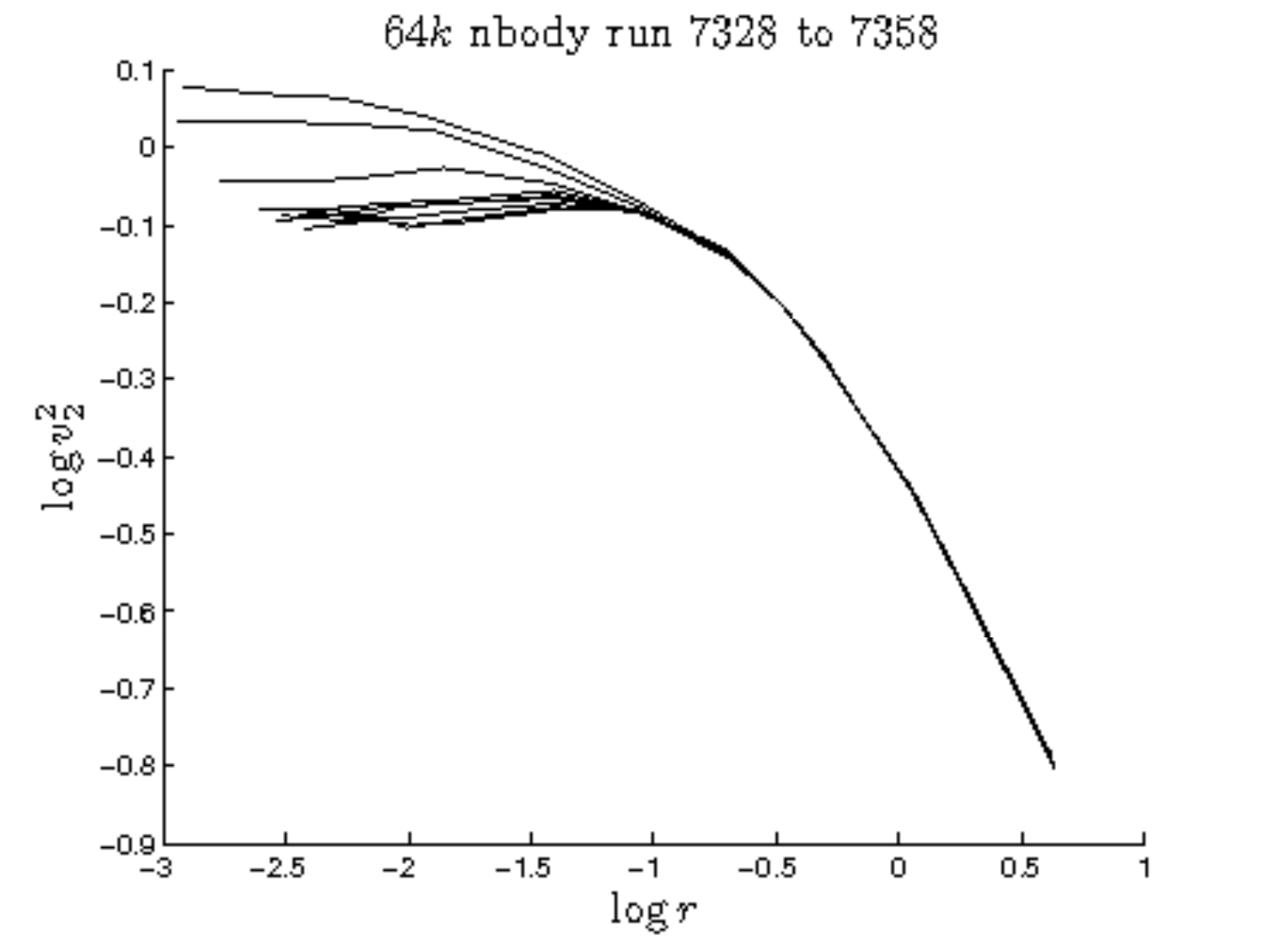}}}

\caption{Top: evolution of the relative binding energy $\frac{E_b}{E}$ and $\log{r_c}$. Bottom: $\log{v_2^2}$ measured at Lagrangian shells in the heavier component over part of the expansion of the core.}
\label{fig:64kb}
\end{figure} 

\begin{figure}
\subfigure{\scalebox{0.55}{\includegraphics{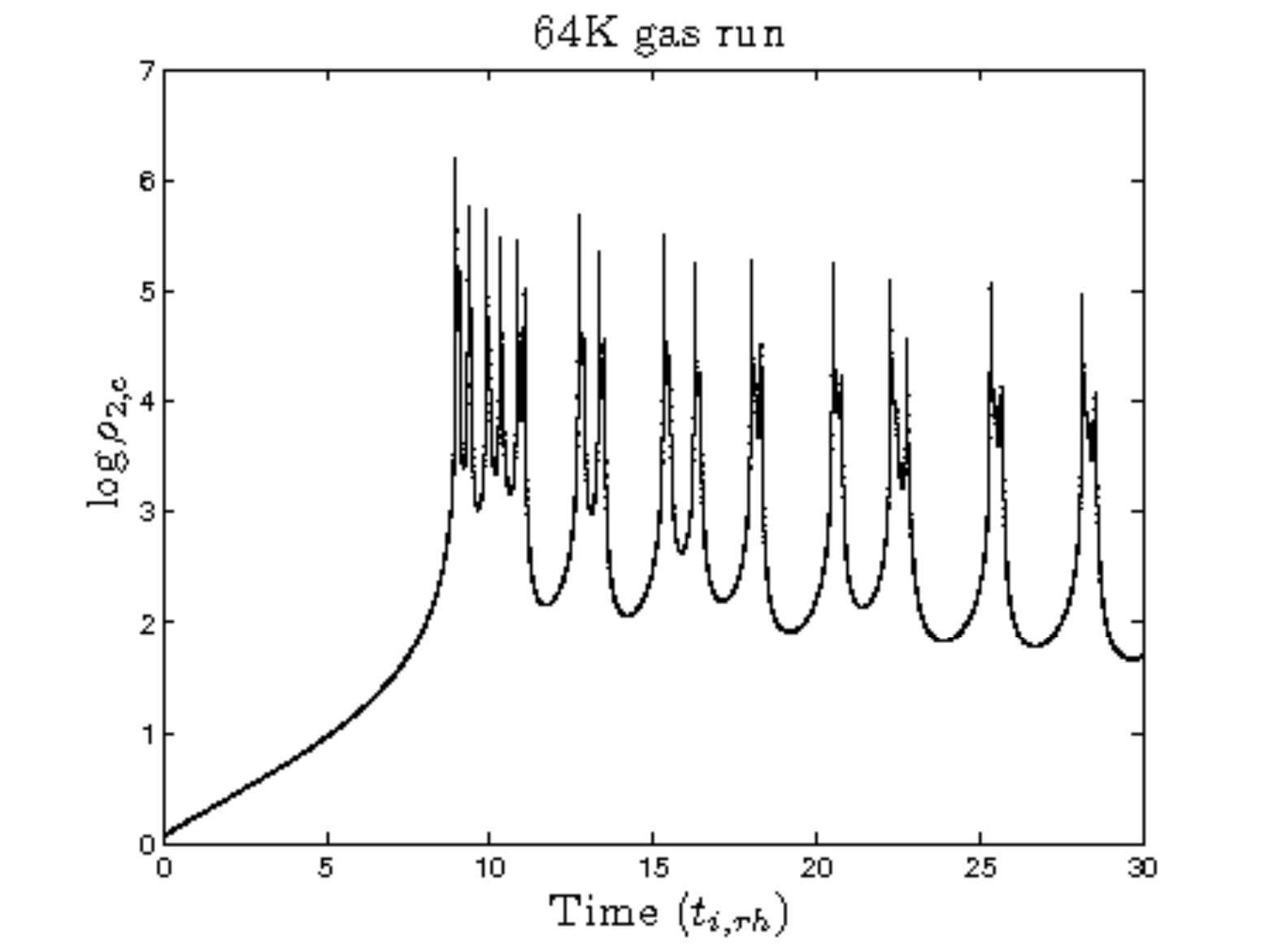}}}\quad
\subfigure{\scalebox{0.55}{\includegraphics{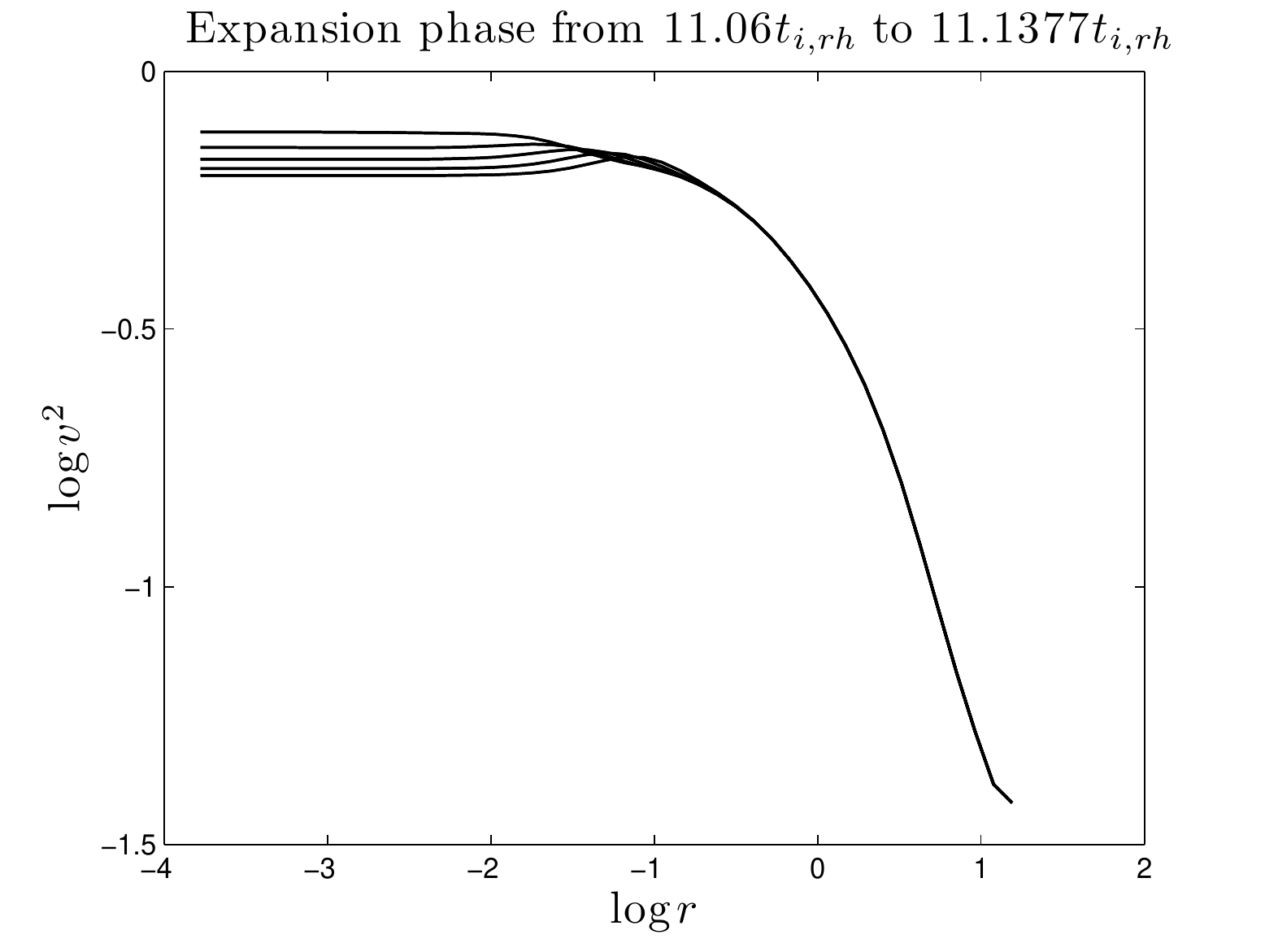}}}
\caption{64k gas model.  Top: $\log{\rho_{2,c}}$ vs time $(units$
  $t_{i,rh})$, bottom: $\log{v_{2}^2}$ vs $\log{r}$ at different
  times during a gravothermal expansion. }
\label{fig:64kgas}
\end{figure} 

The $64k$ run shown in Figs. \ref{fig:64k} and  \ref{fig:64kb} has
large amplitude oscillations. There is a part of the  expansion which
is shown in Fig. \ref{fig:64kb} between $7353$ and $7390$ in which the
relative binding energy of binaries is nearly constant.  Therefore
binary activity cannot be what is driving the
expansion. Fig. \ref{fig:64kb} (bottom) shows the evolution of the profile of $\log{v^2}$ over part of the expansion. A negative temperature gradient is visible towards the end of this expansion and this is what is driving the expansion. From the results of the $32k$ and $64k$ runs it seems that the value of $N_{crit}=17000$  obtained by the gas code is a reasonable indicator of stability for the \Nbody case in the sense that none of the signs of gravothermal behaviour were found for $N \lesssim16k$.

\section{Conclusions and Discussion}

The main focus of this paper has been on the gravothermal oscillations
of two-component systems. The critical value of $N$ for the onset of
instability has been found for a range of stellar mass ratios and
total mass ratios using a gas model. The case of $\frac{M_2}{M_1}=1$
and $\frac{m_2}{m_1}=2$ was further investigated using the direct
\nbody code NBODY6. The value of $N_{crit}$ obtained from the gas code
seems to be a good indicator for stability in \Nbody runs for this
case. Based on this, it is a reasonable assumption that the other
$N_{crit}$ values would give an indication of the stability for direct
\nbody systems. The values of $N_{crit}$ for the two-component model
were found to be much higher than for the one-component case and were
found to vary with $\frac{m_2}{m_1}$ and $\frac{M_2}{M_1}$. However,
the value of $N_2$ at the stability limit was found to vary much less
than $N$ itself. This seems to suggest that instability depends on the
properties of the heavy component (see \ref{sec:lsm}). A possible
explanation of this is given in Section \ref{sec:lsm}. 

The physical manifestation of the oscillations was investigated for the case of small-amplitude periodic oscillations in the gas model. It has been pointed out that the collapse of the region between $r_c$ and $r_h$ is an important mechanism which can halt the expansion phase of a gravothermal oscillation. This mechanism should also be present in one-component models and it would be an interesting topic for future work to see how this mechanism would behave with  different stellar mass functions. 

\cite{KimLeeGood1998} argued that two-component clusters may be
realistic approximations of multi-component clusters, where the two
components are neutron stars and main sequence stars and the effect of
white dwarfs (heavier than the turnoff mass) was assumed to be
negligible. They also only studied cases that were Spitzer stable,
which means that the components were able to achieve equipartition of
kinetic energy. For the two-component case, it is only possible for it
to be Spitzer stable if there is only a small amount of the heavier
component present. As there is a significant range of stellar masses in a real star cluster, it is likely that some form of the Spitzer instability will be present.

To apply our ideas to a multi-component system, it may be possible to
group the heavier components together if they are able to achieve
approximate thermal equilibrium. This could be considered as a single
heavier component which is Spitzer unstable with respect to the remaining components. This would help to reduce a multi-component system to the two-component case studied in this paper.

Nevertheless, it is not clear quantitatively how the considerations of this research are to be applied to a multi-component cluster. Furthermore, we have ignored many things such as primordial binaries, tidal fields and stellar evolution and these are important in the evolution of a real star cluster. Further study is needed in order to understand the phenomenon that is gravothermal oscillation.

\section*{Acknowledgements}
We are indebted to S. Aarseth and K. Nitadori for making publicly available their version of NBODY6 adapted for use with a GPU. We thank the anonymous referee for helping us to clarify the paper. Our hardware was purchased using a Small Project Grant awarded to DCH and Dr M. Ruffert (School of Mathematics) by the University of Edinburgh Development Trust, and we are most grateful for it. PGB is funded by the Science and Technology Facilities Council (STFC).

\bsp

\label{lastpage}

\end{document}